\documentclass[structabstract]{aa}
\usepackage{graphicx}
\usepackage{txfonts}
\begin{document}
   \title{The ionization structure of multiple shell planetary nebulae}

   \subtitle{I. NGC 2438 \thanks{
   Based on observations at ESO and SAAO, and HST archival data}}

   \author{S. {\"O}ttl
          \inst{\ref{inst1}}
          \and
          S. Kimeswenger\inst{\ref{inst2},\ref{inst1}}
          \and
          A. A. Zijlstra\inst{\ref{inst3}}
          }

   \institute{Institute for Astro and Particle Physics, Leopold Franzens Universit\"at Innsbruck,
              Technikerstrasse 25, A-6020 Innsbruck, Austria\\
              \email{Silvia.Oettl@uibk.ac.at}\label{inst1}
   \and
   Instituto de Astronom{\'{i}}a, Universidad Cat{\'{o}}lica del Norte,
                Avenida Angamos 0610, Antofagasta, Chile\newline
              \email{Stefan.Kimeswenger@gmail.com}\label{inst2}
   \and
   Jodrell Bank Centre for Astrophysics, School of Physics and Astronomy, University of Manchester, Manchester M13 9PL, UK\newline\email{a.zijlstra@manchester.ac.uk}\label{inst3}
             }

   \date{Received December 06, 2013; accepted xxxx nn, 2014}

% \abstract{}{}{}{}{}
% 5 {} token are mandatory

\abstract{In recent times an increasing number of extended haloes and multiple shells
  around planetary nebulae have been discovered. These faint extensions to the main
  nebula
  %(called the 'rim')
  trace the mass-loss history of
  the star, modified by the subsequent evolution of the nebula.  Integrated
  models predict that some haloes may be recombining, and not in ionization
  equilibrium. But parameters such as the ionization state and thus the
  contiguous excitation process are not well known. The haloes are
  very extended, but faint in surface brightness - $10^3$ times below the main
  nebula. The observational limits lead to the need of an extremely well
  studied main nebula, to model the processes in the shells and
  haloes of one object. NGC~2438 is a perfect candidate to explore the physical
  characteristics of the halo.  }
   {The aim is to derive a complete data set of the main nebula. %rim.
    This allows us to derive the physical conditions, such as temperature, density and ionization, and clumping, from photoionization
    models. These models are used to derive whether the halo is in ionization
    equilibrium.}
 {Long-slit spectroscopic data at various
   positions in the nebula were obtained at the ESO~3.6\,m and the SAAO~1.9\,m
   telescope. These data are supplemented by imaging data from the HST archive and from the ESO~3.6\,m telescope, and
   archival VLA observations. The use of diagnostic diagrams draws limits for
   physical properties in the models. The photoionization code CLOUDY is used
   to model the nebular properties, and to derive a more accurate
   distance and ionized mass.}
  {We derive an accurate extinction $E_{\rm B-V}=0.16$, and distance of
   $1.9\pm0.2$kpc. This puts the nebula behind the nearby open cluster M46 and
   rules out membership. The low-excitation species are found to be dominated
   by clumps. The emission line ratios show no evidence for shocks. The
   filling factor increases with radius in the nebula. The electron densities
   in the main nebula are $\sim 250$cm$^{-3}$,
   dropping to $\sim 10$--30 in the shell.  We find the shell in
   ionization equilibrium: a significant amount of UV radiation infiltrates the
   inner nebula. Thus the shell still seems to be ionized. The spatially
   resolved CLOUDY model supports the hypothesis that photoionization is the
   dominant process in this nebula, far out into the shell. Previous models
   predicted that the shell would be recombining, but this is not confirmed by
   the data. We note that these models used a smaller distance, and therefore
   different input parameters, than derived by us.
  }
   {}


   \maketitle
%
%________________________________________________________________
%\def\col_version{(The color version of the graph is provided in the online journal.)}

\section{Introduction}

{\it Multiple shell planetary nebulae (MSPNe):}
\newline
Planetary nebulae (PNe) are the ionized ejecta from an asymptotic giant branch
(AGB) star.  They are a short-lived phenomenon compared to a typical stellar
lifetime, and are visible while the now post-AGB star crosses the
Hertzsprung-Russell diagram (HRD) towards high temperatures, before entering
the white-dwarf cooling track. Most of the luminous material originates from
the stellar wind during the last thermal pulses on the AGB.

This paper deals with the physical conditions of a special type of PNe:
Multiple-shell planetary nebulae (MSPNe), which are surrounded by faint outer
shells and/or haloes. In recent times, an increasing number of haloes and
multiple shells around PNe have been discovered. Although MSPNe are a familiar
phenomenon, most of them are poorly studied and understood. Previous research
(e.g., Corradi et al. \cite{corradi_03}, Zhang et al. \cite{zhang12} or
Ramos-Larios et al. \cite{IC418}\&\cite{NGC6369}) identified the existence as
common feature for nearly-round nebulae. They appear during the evolution
around the knee in the HRD and in the early part of the cooling track. The
observed MSPN structures are the intricate result of the interaction of
hydrodynamic and radiative processes, both during the AGB and the post-AGB
phases. A detailed description of the mass loss history and the connection
between the stellar winds and the huge extended circumstellar envelopes can be
found in, e.g., Bl{\"o}cker (\cite{bloecker_95}) and Decin (\cite{decin}). Due
to the faintness of the haloes, up to a factor of $10^3$ below the main nebula
in surface brightness, most of them were discovered much later than the nebula
itself.

Corradi et al. (\cite{corradi_00}), Sch\"{o}nberner \& Steffen
(\cite{schoenberner_02}) and Perinotto et al. (\cite{perinotto_04}) used 1D
radiative transfer hydrodynamic (RTH) models to calculate the evolution of
this kind of nebulae. They modeled the full evolution of the PN starting at
the AGB in a sophisticated way. These important models provide a very good
representation of the surface brightness of MSPNe at an age of about 10\,000
years. The RTH models result in outgoing shock fronts. At an age of about
10\,000 years (assuming the Bl{\"o}cker (\cite{bloecker_95}) track of an
0.605\,M$_\odot$ central star), the resulting density profile peaks at $n_{\rm
  H}\gtrapprox$ 500~cm$^{-3}$ (and thus $n_{\rm e}\gtrapprox$
560~cm$^{-3}$). The central star of the PN (CSPN) has a luminosity of
$\approx$250~$L_\odot$ and a temperature of $T_{\rm CSPN} \approx 120$~kK. In
those models, the outer limit of the bright nebula is the limit of the hard UV
radiation -- a {\sl radiation bounded} optically thick PN. The shell material
is nearly recombined ($n_{\rm e} : n_{\rm H} = 1 : 10$), with a temperature of
2\,000~K. Thus the [\ion{O}{i}] $\lambda$6300\AA~ line is predicted to be
about 6 times stronger than \ion{H}{$\beta$}.

These kind of models are enormously important for the general understanding of
the evolution. The layered structure predicted by the models compares well
with observations. In this paper, we will investigate the ionization structure
of a MSPN, in order to provide further observational constraints for the
models.

In this work we use the nomenclature introduced by Chu et al. (\cite{chu_87}) and Balick et al. (\cite{balick_92}), but fully defined by Corradi et al. (\cite{corradi_03}), based on the evolutionary RTH models:
\begin{itemize}
\item{}the main nebula - also called 'rim' by them;
\item{}the first thin surrounding structure including its outer weak intensity enhancement is called 'shell';
\item{} the faint outer structures are called 'halo1' and 'halo2'.

\end{itemize}

\smallskip
\noindent{\it NGC 2438:}\newline
Our target is a classical MSPN, used as 'benchmark' for the
modeling (Corradi et al. \cite{corradi_00}). NGC 2438 shows a bright inner
main nebula; the geometry of the nebula is near round and closed. The diameter
of the main nebula is about 60\arcsec. The nebula indicates two slightly detached
shells and a very faint halo (Fig. \ref{slits.fig}). This faint halo is most
visible in the western part of the nebula and seems to have a circular
shape. We can find ray-like structures and clumps in the nebula as well. The
CSPN is not the bright star near the center, but the fainter one at the
center of the nebula.

The only observational study of the nebula over a wide optical wavelength
range and through the whole main nebula was obtained by Guerrero \& Manchado
(\cite{gu_ma_99}), with a single exposure at the ESO 1.5m telescope. They
centered the list on
the brightest star near the center (not the CSPN), and
integrated over wide areas along the slit. Guerrero \& Manchado
(\cite{gu_ma_99}) report no detection of the [\ion{O}{i}] line. This might be
due to the strength of the telluric airglow line at this
position. Further studies of the innermost regions around the CSPN were given
by Torres-Peimbert \& Peimbert (\cite{ToPe77}), Kaler (\cite{Ka83}),
Kingsburgh \& Barlow (\cite{KB94}) and Kaler et al. (\cite{Ka90}). All of them
were focusing on abundances and they find a mild helium and nitrogen
overabundance.\newline
The spectral investigation of Corradi et al. (\cite{corradi_00}) covered the
regions of [\ion{O}{iii}] $\lambda5007$\AA~ and \ion{H}{$\alpha$} +
[\ion{N}{ii}] $\lambda$6548\AA+$\lambda$6584\AA~ with a  high resolution
of $n 70\,000$. They provide good results for the expansion of the nebula.
\newline
The CSPN was investigated in detail by Rauch et al. (\cite{rauch_99}), using n
on-local thermodynamic equilibrium (NLTE) stellar atmosphere models. The
results are $\log(g[\rm cgs]) = 6.62\pm0.22$, $T_{\rm CSPN} = 114\pm10\,$kK,
$L_{\rm CSPN} = 570\,L_\odot$, and $M_{\rm CSPN} =
0.56\pm0.01\,\,M_\odot$. The helium overabundance of the CSPN is slightly
above the one found in the studies of the inner nebula mentioned before. Rauch
et al. (\cite{rauch_99}) report that the nebula luminosity is an order of
magnitude above the luminosity of the CSPN. We later show (see
Sec. \ref{distance.sec}), that this discrepancy was not caused by the model,
but by the photometry from literature they used. The low CSPN mass would imply
a slow post-AGB evolution.
\newline
Based on the line ratios of [\ion{O}{iii}] : \ion{He}{ii} : \ion{H}{i} in the
main nebula and the line ratio of [\ion{O}{iii}] : \ion{H}{i} in the shell,
photoionization studies (Armsdorfer et al. \cite{armsdorfer_02},
\cite{armsdorfer_03}) state that the shell consists of ionized material. The
required amount of ionizing UV photons can be obtained by a clumpy structure of
the main nebula, allowing UV photons to escape. Such structures are
established for some well-studied PNe, e.g., the Helix Nebula (O'Dell et
al. \cite{helix_05}; Matsuura et al. \cite{helix_09}) or the Ring Nebula
(Speck et al. \cite{ring_03a}; O'Dell et al. \cite{ring_03b}). In Dalnodar \&
Kimeswenger (\cite{apn5}), the positions of the spokes of enhanced intensity
in the shell of NGC 2438 was shown to be correlated to holes in the main nebula. This
supports the
concept of a {\sl matter bounded} structure. Recent studies of the MSPN IC~418
(Ramos-Larios et al. \cite{IC418}) and NGC~6369 (Ramos-Larios et
al. \cite{NGC6369}) reveal very similar results: {\sl 'Radial filaments
  emanate outwards from most of the }[\ion{N}{ii}]~{\sl knots'}.

\medskip
In this work we compiled own spectroscopy and multi-wavelength imaging data
sets. Combining more information and data gives us the opportunity to search
for the origin of the reported discrepancies and to draw detailed constraints
for multi-dimensional RTH studies of MSPNe. We analyze long slit
spectra, narrow-band images, a VLA radio map and HST archival data. We
investigate the main nebula and the shell by means of a sophisticated spatially
resolved CLOUDY model (Ferland et al. \cite{cloudy}, \cite{cloudy13}).

\section{Observations and data reduction}

\subsection{Spectroscopy}
 \begin{figure}[!ht]
   \centering
   \includegraphics[bb=0.1 0.1 8.8cm 8.38cm, width=8.8cm, clip=true]{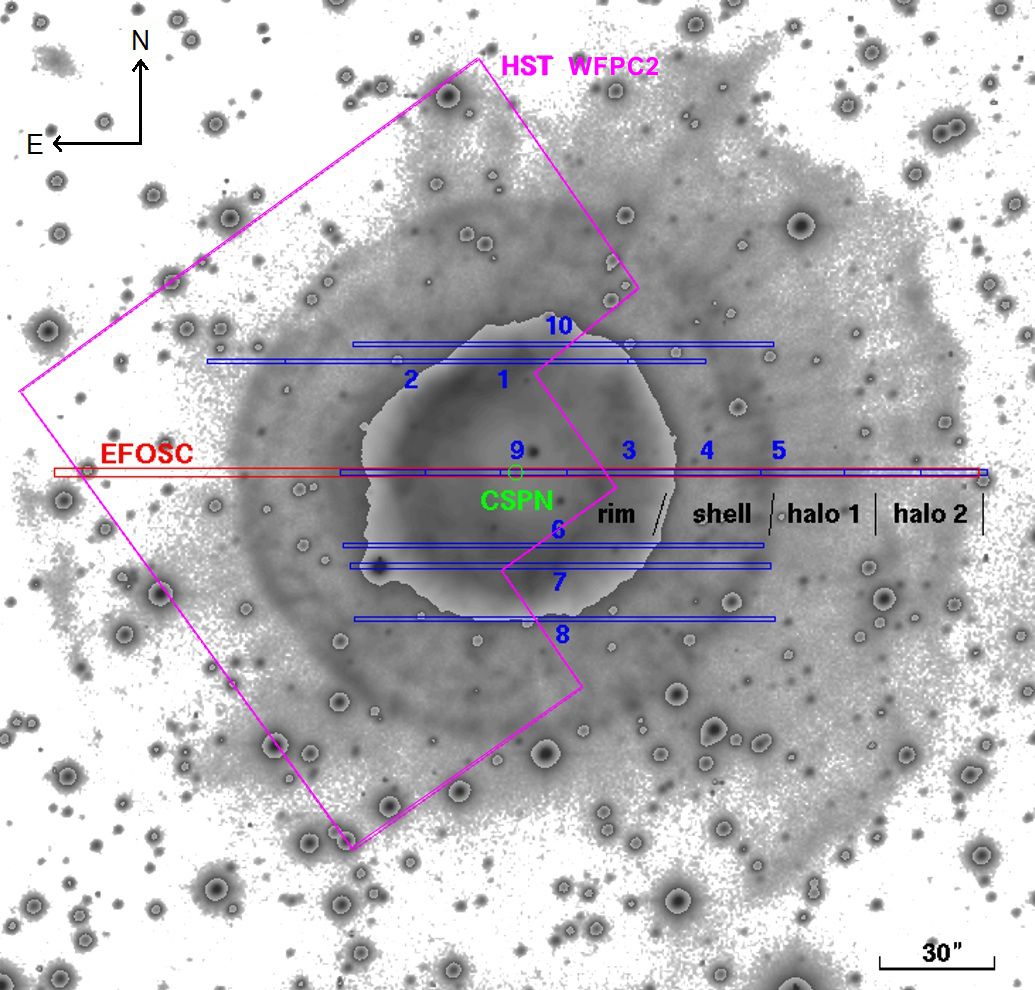}
      \caption{An overview of the target observations. The background image is
        a \ion{H}{$\alpha$} EFOSC1 image. The spectrograph slits were all
        positioned in the E-W direction. The SAAO slits numbers correspond to
        those in Tab.~\ref{spec}. The CSPN and the region names as defined
        in Sec.~1 are shown and the field of view of the HST WFPC2 images is
        indicated as well.}
         \label{slits.fig}
   \end{figure}

Two independent spectroscopic data sets of NGC 2438 were used. The first
spectroscopic data set was taken at the European Southern Observatory (ESO) at
La Silla in Chile. Observations during three nights in 1996 were obtained,
using the 3.6~m telescope and the ESO Faint Object Spectrograph and Camera
(EFOSC1\footnote{www.eso.org/sci/facilities/lasilla/instruments/efosc/History.html})
in long slit mode. The slit was positioned across the CSPN. These
observations are summarized in Tab.\ref{spec}.

\noindent The second spectroscopic data set was taken 2002 at the South
African Astronomical Observatory (SAAO) 1.9~m Radcliffe Telescope in
Sutherland. The observations were obtained during four nights. Due to the
short slit, several positions were taken - scanning the whole nebula. These
observations are summarized in Tab.~\ref{spec}. The slit positions are
indicated in Fig.\,\ref{slits.fig}.

\noindent All instrumental parameters of both observations and all the technical details are summarized ind Tab.~\ref{tech}.

\begin{table}[!ht]
\caption{The spectroscopic data set. }
\label{spec}
\centering
\begin{tabular}{cccccc}
\hline\hline
 & Position\ \ \ \ \ \ \ \ \ \ \ \phantom{x} &Date UT & airm. & expo.\\
Nr. & RA\ \ \ \ \ \ \ \ \ \ \ DEC & & &[sec]\\
\hline
 & {\sl ESO data}\ \ \ \ \ \ \ \ \ \ \ \phantom{x} & & &\\
\hline
&7$^{\rm h}$41$^{\rm m}$50\fs51\ \ -14\degr44\arcmin07\farcs7 & 02-11~02:52&1.034&3600\\
&7\phantom{$^{\rm h}$}41\phantom{$^{\rm m}$}50.51\ \ -14\,\,44\,\,07.7 & 02-13~01:18 &1.119 &3600\\
\hline
 & {\sl SAAO data}\ \ \ \ \ \ \ \ \ \ \ \phantom{x} & & &\\
\hline
1a&7$^{\rm h}$41$^{\rm m}$50\fs88\ \ -14\degr43\arcmin38\farcs5 & 04-16~17:31&1.01&1800\\
1b&7\phantom{$^{\rm h}$}41\phantom{$^{\rm m}$}50.87\ \ -14\,\,43\,\,38.5 &04-16~18:04&1.07&600\\
2&7\phantom{$^{\rm h}$}41\phantom{$^{\rm m}$}52.22\ \ -14\,\,43\,\,38.5 &04-16~18:19&1.08&1800\\
3&7\phantom{$^{\rm h}$}41\phantom{$^{\rm m}$}48.44\ \ -14\,\,44\,\,07.7 &04-17~17:21&1.01&1800\\
4&7\phantom{$^{\rm h}$}41\phantom{$^{\rm m}$}47.13\ \ -14\,\,44\,\,07.7 &04-17~18:18&1.08&1800\\
5&7\phantom{$^{\rm h}$}41\phantom{$^{\rm m}$}45.96\ \ -14\,\,44\,\,07.7 &04-17~18:53&1.08&1800\\
6&7\phantom{$^{\rm h}$}41\phantom{$^{\rm m}$}49.85\ \ -14\,\,44\,\,27.0 &04-18~17:43&1.01&1800\\
7&7\phantom{$^{\rm h}$}41\phantom{$^{\rm m}$}49.73\ \ -14\,\,44\,\,32.3 &04-18~18:19&1.08&1800\\
8&7\phantom{$^{\rm h}$}41\phantom{$^{\rm m}$}49.67\ \ -14\,\,44\,\,46.3 &04-18~18:52&1.25&1800\\
9&7\phantom{$^{\rm h}$}41\phantom{$^{\rm m}$}49.91\ \ -14\,\,44\,\,07.7 &04-19~17:38&1.03&1800\\
10&7\phantom{$^{\rm h}$}41\phantom{$^{\rm m}$}49.68\ \ -14\,\,43\,\,34.0 &04-19~18:42&1.12&3600\\
\hline
\end{tabular}
\tablefoot{In the ESO observations, the slit was centered and guided at the CSPN. In the SAAO observations, the slit was moved to various positions in the target, indicated by the numbers. The coordinates were calibrated later using 2MASS stars. The airmass in both observations is given at mean time of the observation.}
\end{table}

\begin{table}[!ht]
\caption{Technical details and instrumental parameters.}
\label{tech}
\centering
\begin{tabular}{ccc}
\hline\hline
Observatory & ESO & SAAO\\
\hline
CCD & \# 26 Textronic & SITe \\
pixel & 571 $\times$ 520 & 266 $\times$ 1798 \\
size & 27 micron & 15 micron\\
grism/grating & \# 10 - 150 lines/mm & \# 7 - 300 lines/mm \\
slit width & 1\arcsec & 1\arcsec \\
slit length & 243\arcsec & 110\arcsec \\
direction & E-W fixed & E-W \\
spatial separation & 0\farcs64 per pixel & 0\farcs73 per pixel\\
wavelength range & $\lambda$3780\AA\, to \,$\lambda$5508\AA\ & $\lambda$3700\AA\, to\,$\lambda$7400\AA\ \\
sampling & 3.0\,\AA~ per pixel & 2.3\,\AA~ per pixel \\
resolution & 6.0\,\AA & 4.5\,\AA \\
\hline
\end{tabular}
\tablefoot{The spatial separation is along the slit. The blaze wavelength for the SAAO grating \# 7 is 4600\AA . }
\end{table}

\begin{figure}[!th]

\begin{tabular}{ll}
$\!\!$\includegraphics[bb=.2cm 0 13.55cm 13.55cm,clip, width=4.349cm]{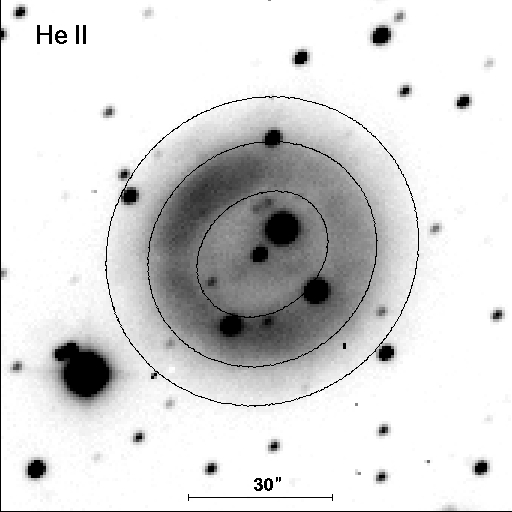}&$\!\!\!\!\!\!$\includegraphics[bb=.2cm 0 13.55cm 13.55cm,clip, width=4.349cm]{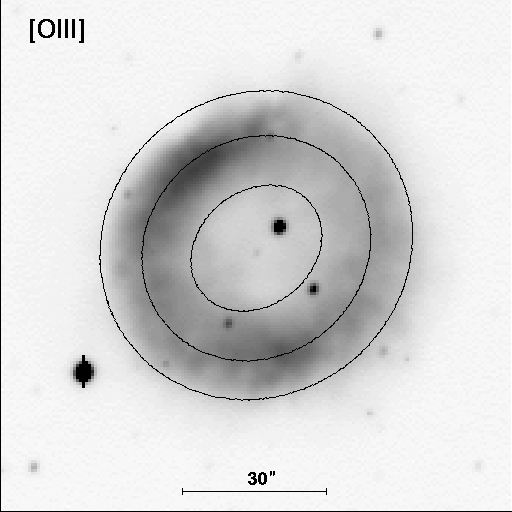}\\
$\!\!$\includegraphics[bb=.2cm 0 13.55cm 13.55cm,clip, width=4.349cm]{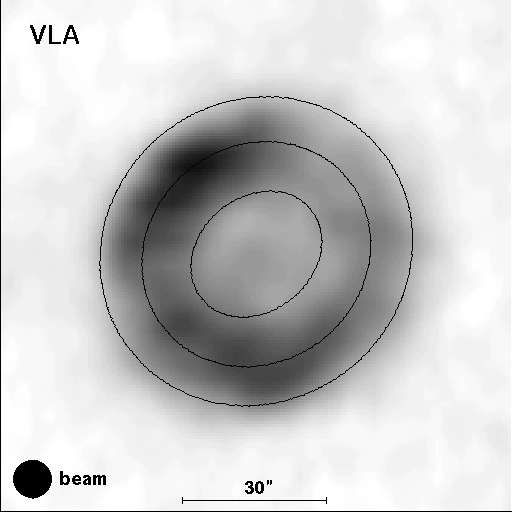}&$\!\!\!\!\!\!$\includegraphics[bb=.2cm 0 13.55cm 13.55cm,clip, width=4.349cm]{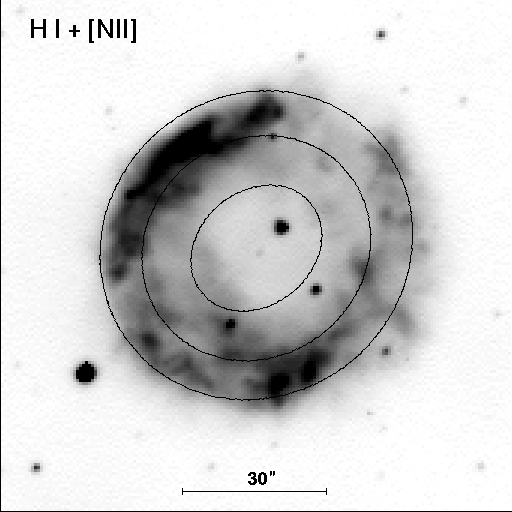}\\
\end{tabular}
\caption{The direct images of the main nebula. First row: \ion{He}{ii} ($\lambda$4686\AA) \& [\ion{O}{iii}] ($\lambda$5007\AA); second row: VLA (20 cm) \& \ion{H}{$\alpha$} ($\lambda$6563\AA) +[\ion{N}{ii}] ($\lambda$6548\AA+$\lambda$6584\AA) (same orientation as shown in Fig.~\ref{slits.fig} - N is up, E is left).
The ellipses are derived from the images of the \ion{He}{ii} line. While the \ion{He}{ii} emission is concentrated clearly towards the inner edge of the main nebula, the [\ion{O}{iii}] and VLA images have about the same appearance.
} \label{imaging}
\end{figure}

The spectra were reduced by standard MIDAS (Warmels \cite{midas})
routines. Careful corrections for variations of the slit width along the
  slit were applied by adding the flatfield images. The variations were
  about 2\% in the EFOSC1 spectra and up to 17\% in the SAAO spectra. The
  variations show no dependency on wavelength. As the outer haloes are
not visible east of the nebula in the ESO data, the night sky could be taken
directly from the spectra. For the SAAO data, the sky was taken from the slit
positions outside the nebula and the shell. Additionally, the sky for the SAAO
data could be taken from data from other, smaller nebulae, which were observed
by Thomas Rauch during the same nights. These other PNe without halo were
  used to estimate straylight contamination of the surroundings.  Both
  instruments do not show any contamination down to the noise level of the
  images. For the extraction of the nebular lines and proper suppression of
  local free-free continuum, the nearest empty region along the slit was
  used. Usually, the nearest region was directly redwards and bluewards of
  each line. In some cases only one side was used, to avoid blending with
  other lines. To improve S/N, the width of the region was 3 times the FWHM of
  the lines. The contribution was found to be fairly small within the main nebula.
  Only at the very blue end the readout noise was detected a few times.
  Outside the main nebula, towards shell and halo, no variation above the sky level
  within the limits of the readout noise was detected.

\begin{figure*}[!th]
\sidecaption
{\begin{tabular}{ll}
$\!\!$\includegraphics[bb=0 0 5.9cm 5.9cm, width=5.9cm]{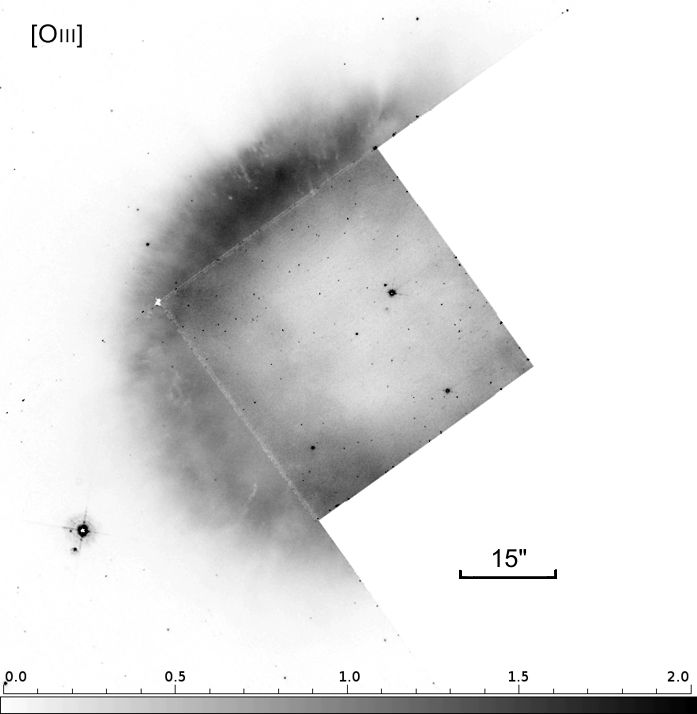}&$\!$\includegraphics[bb=0 0 5.9cm 5.9cm, width=5.9cm]{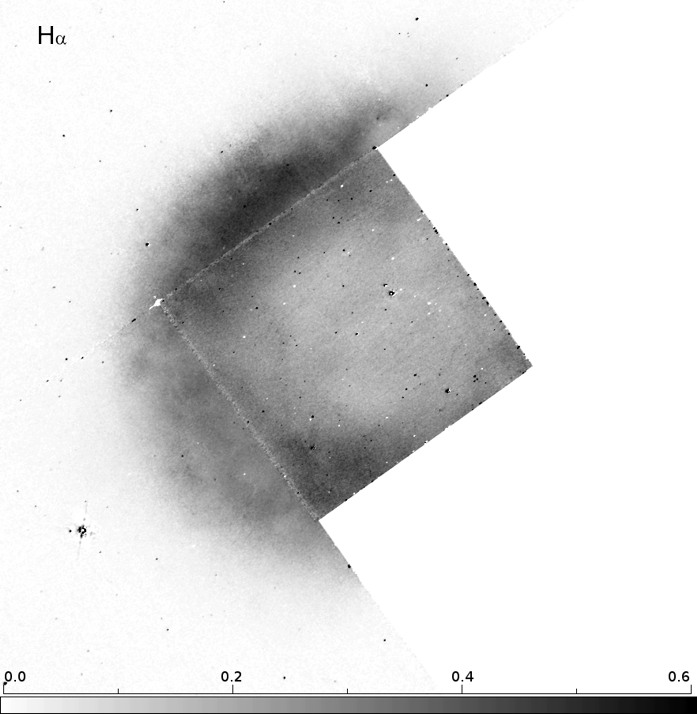}\\
$\!\!$\includegraphics[bb=0 0 5.9cm 5.9cm, width=5.9cm]{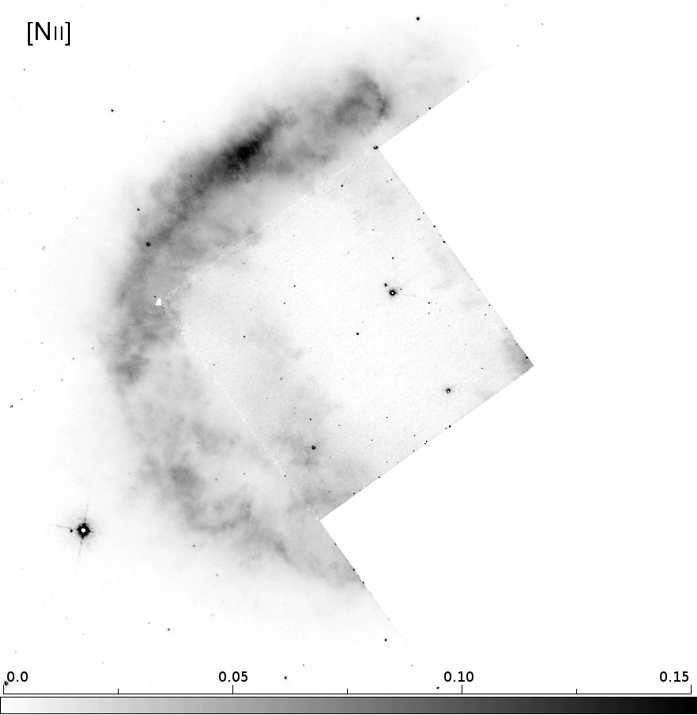}&$\!$\includegraphics[bb=0 0 5.9cm 5.9cm, width=5.9cm]{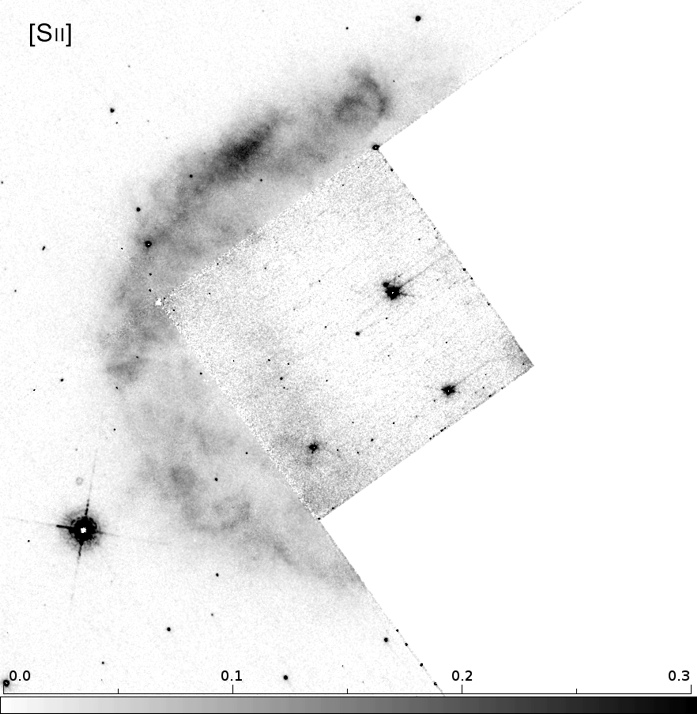}\\
\end{tabular}
}
\caption{The HST images of the main nebula. First row: [\ion{O}{iii}]($\lambda$5007\AA)
 and \ion{H}{$\alpha$} after subtracting [NII] contribution (see text); second row: [\ion{N}{ii}] ($\lambda$6584\AA) and [\ion{S}{ii}]($\lambda$6716\AA+$\lambda$6732\AA). The images are in the same orientation as shown in Fig.~\ref{slits.fig} and cover the field of view shown there. \newline The similarities in the structures are obvious.}
 \label{imaging_hst}
\end{figure*}

Corrections of differential refraction were applied on both data sets, using the algorithm of Fluks \& The
(\cite{fluks_the_92}). These corrections were applied to the standard
stars as well as to the CSPN spectra, to obtain the absolute flux calibrations.
Due to the observations near zenith, the corrections for the nebula
are very small and the effect of variation of the parallactic angle during the
long nebula exposures are negligible. In case of the standard stars, we have
exposures of 1 and 5 minutes only, during which the parallactic angle was
constant. The differential refraction correction is up to 9\% at the blue end
of the spectra.

After applying all corrections from night to night, the flux differences of
the standard stars were only a few percent. For overlapping observations
between the SAAO and the ESO spectra, a variation below 5~\% of the bright
lines (e.g., [\ion{O}{iii}]) was found. For the fainter lines (e.g.,
\ion{Ar}{ii}, \ion{H}{$\delta$}, \dots), a typical variation of less than
10~\% was found. Taking conservative estimates of systematic effects, we
assume the overall absolute flux accuracy to be better than 12~\% in the main
nebula, and better than 20~\% in the shell. Differential values (e.g., line
ratios) are expected to be better than 5~\%.

\subsection{Imaging}
\label{imaging.sec}
Direct images using narrow-band filter [\ion{O}{iii}] $\lambda5007$\AA~ (ESO
filter \#686), \ion{He}{ii} $\lambda4686$\AA~ (ESO filter \#512) and
\ion{H}{$\alpha$} + [\ion{N}{ii}] $\lambda$6548\AA~/$\lambda$6584\AA~) (ESO
filter \#691) were obtained at the ESO 3.6 m telescope. The absolute flux
calibration was scaled from the long slit spectra. The \ion{H}{$\alpha$}
filter \#691 contains significant contributions of both [\ion{N}{ii}]
lines. The filter response was used to derive a sensitivity of 95\% of the
$\lambda$6548\AA~ line and 75\% of the $\lambda$6583\AA~ line (relative to
peak transmission).

\noindent As a byproduct of an investigation of the nearby bipolar nebula
OH~231.8+4.2 (Taylor \& Morris \cite{vla}), VLA radio observations of NGC 2438
were obtained and made available to us by the authors.

The morphologies in the different images show variations
(Fig.~\ref{imaging}). In the [\ion{O}{iii}] image and in the VLA image, the
same nebula structures, elongation, and central cavity were found. The
\ion{He}{ii} image shows more concentration towards the inner region. Despite
the Bowen fluorescence, the decoupling of the helium image and the oxygen
image shows the limit of the optically thick region in the helium Lyman
continuum, used in the modeling part (Sec.\,\ref{model.sec}). The
(\ion{H}{$\alpha$}$+$[\ion{N}{ii}]) image is dominated by substructures and
clumps. The rim of the main nebula is slightly elongated by about 0.9~:~1
(Fig.~\ref{imaging}). The ellipticity is fairly small (0.84, 0.91 and 0.94
from inner to outer ellipse) and the position angles are constant.

\begin{figure}[!th]
\centering
\includegraphics[bb=0 0 8.8cm 10.8cm, width=8.0cm]{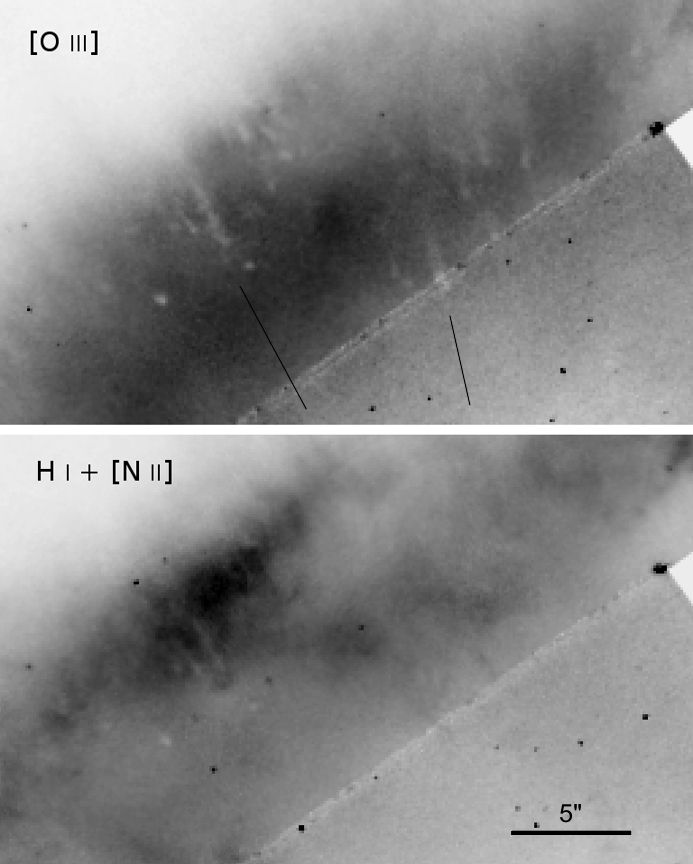}
\caption{A zoom of the N-E region of the main nebula from the HST images.
The cometary knots are especially pronounced in the [\ion{O}{iii}] images.
The shadows point exactly away from the CSPN (direction indicated by two lines).}
\label{zoom_hst}
\end{figure}

Additionally, a data set of NGC 2438 in the HST archive was retrieved. The
images, taken in September 2008, do not cover the whole nebula. The data set
is summarized in Tab.~\ref{hst_tab}. The inner region was covered by the high
resolution PC camera and has a very low signal to noise ratio. The images are
not deep enough to show  structures of the shell or the haloes
(Fig.~\ref{imaging_hst}). They were used to analyze shock signatures (see
Sec.~\ref{shock.sec}).

\begin{table}[!ht]
\caption{Observation log of the HST archival images (all taken with WFPC2).}
\label{hst_tab}
\centering
\begin{tabular}{c c c c}
\hline\hline
 line & filter & exposure time & observing  \\
      &        &  time [s]   &  date\&time\\
\hline
\ion{H}{${\alpha}$}+[\ion{N}{ii}] & F656N & 1000 & 2008-09-19 09:51:17\\

 $[$\ion{O}{iii}$]$\phantom{XXI}& F502N & 1000 & 2008-09-19 11:25:17\\
 $[$\ion{N}{ii}$]$\phantom{XiiII} & F658N & \phantom{1}500 &2008-09-19 13:01:17\\
 $[$\ion{S}{ii}$]$\phantom{XXII} & F673N & \phantom{1}400 &2008-09-19 13:13:17\\	
\hline
\end{tabular}
\end{table}

\noindent The post-flight measurements of the WFPC2 filters, the expansion
velocity and the systematic velocity, measured by Meatheringham et
al. (\cite{vexp_88}) and by Corradi et al. (\cite{corradi_00}), were used to
calculate the effect of the line width on the filter transmission. We obtained
a contamination by [\ion{N}{ii}]$\lambda$6548, and [\ion{N}{ii}]$\lambda$6584
in the F656N filter band of 39\%\ and 2\%, respectively. In contrast, the
F658N filter has a sensitivity of just about 1.7\% for \ion{H}{$\alpha$}. Both
filters have the same peak transmission. Thus we were able to directly obtain
a pure \ion{H}{$\alpha$} image, using the image of F658N and the known line
ratio of [\ion{N}{ii}]$\lambda$6548 / [\ion{N}{ii}]$\lambda$6584. This
\ion{H}{$\alpha$} image shows exactly the same morphology as the
[\ion{O}{iii}] image and VLA image shown before. The [\ion{N}{ii}] image
and the [\ion{S}{ii}] image indicate identical structures throughout all
regions of the nebula. Additionally, the [\ion{O}{iii}] and the
\ion{H}{$\alpha$} images show some cometary knots and ray-like structures
pointing outwards the edges (Fig.~\ref{zoom_hst}). The shadow tails of the knots, pointing
exactly away from the CSPN as result of the shielding of hard UV, are
clearly visible only in [\ion{O}{iii}]. The radiation to ionize the \ion{H}{I} is
provided by the diffuse nebula radiation, thus no shadows are formed in
\ion{H}{$\alpha$}.

\section{Extinction and CSPN parameters}
\label{extinct_and_cspn}
The foreground extinction towards the nebula was derived using the Balmer line
series in the SAAO spectra. The intrinsic line intensities $R_{\rm intr}$ for
the case B recombination calculated by CLOUDY and the interstellar extinction
curve from Osterbrock
\& Ferland (\cite{osterbrock}) of the four isolated Balmer lines from
$\lambda$4102\AA~ to $\lambda$6563\AA~ were used to derive the interstellar
foreground extinction from the measured line ratios $R$ by:
$$ E_{\rm B-V}\,=\,a\,\times\,\log\left({R \over R_{\rm intr}}\right)\,=
\,0\fm16\,\pm\,0\fm01\,\,.$$
\noindent The intrinsic line ratios $R_{\rm intr}$ were determined
  iteratively, starting with $T_{\rm e}\,=\,10$kK, taken from the table in
  Osterbrock \&\ Ferland (\cite{osterbrock}), and calculated with CLOUDY using
  the temperature derived from the \ion{N}{ii} lines (see
  Sec. \ref{model.sec}, Tab. \ref{model_input.tab}).  The final CLOUDY model
  gives additional information about the blended \ion{He}{ii} lines.  The flux
  contribution at these temperatures is 1-3\% only. Thus the error introduced
  for line ratios by neglecting the contribution of the \ion{He}{ii} lines
  should be in the order of 1\% only.  As shown in Tab.\ref{extinct_tab}, all
line ratios lead to the same extinction. Thus the assumptions of case B
recombination and standard extinction apply very well. The spread is
remarkably small.

\begin{table}[!ht]
\caption{Extinction measurement with the Balmer lines.}
\label{extinct_tab}
\centering
\begin{tabular}{c c c c c}
\hline\hline
 line & $R$ & $R_{\rm intr}$ & $a$ & $E_{\rm B-V}$\\
\hline
H$_{\alpha}$/H$_{\beta}$ & 3.300 & 2.790 & \phantom{-}2.21\phantom{0} & 0.1613 \\
H$_{\gamma}$/H$_{\beta}$ & 0.443 & 0.476 & -0.517 & 0.1599 \\
H$_{\delta}$/H$_{\beta}$ & 0.236 & 0.262 & -3.52\phantom{0} & 0.1612 \\
\hline
\end{tabular}
\end{table}

\begin{figure}[!ht]
\centering
\includegraphics[bb=0 0 8.8cm 5.9cm, width=8.8cm]{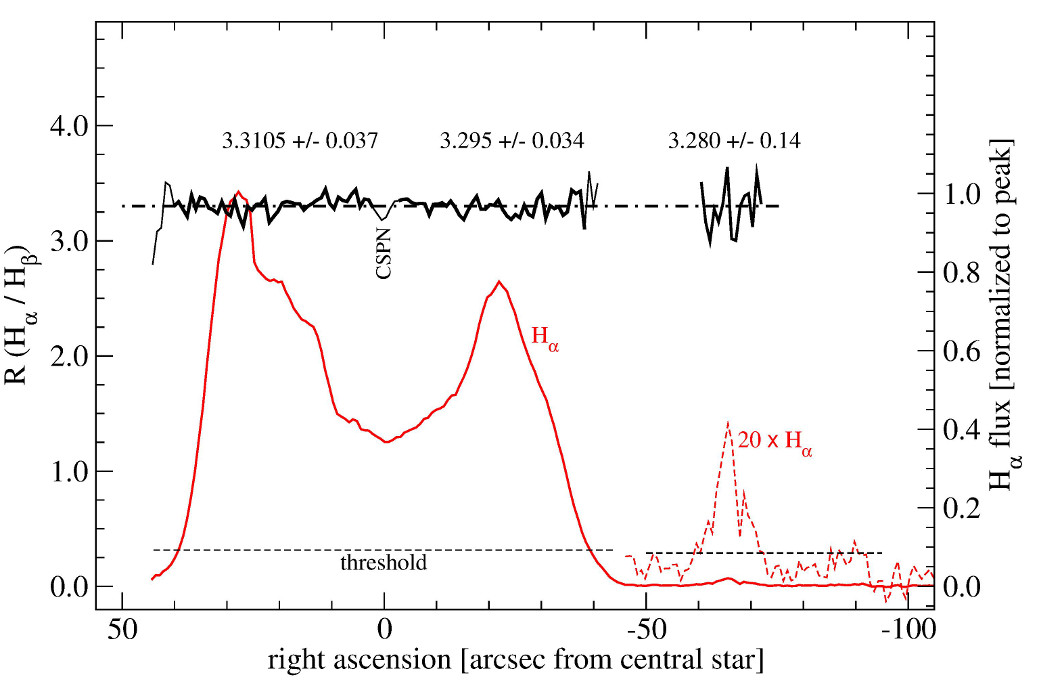}
\caption{The spatial distribution of the line ratio $R = \ion{H}{\alpha}/\ion{H}{\beta}$ to search
for intrinsic extinction. To cover homogeneous S/N ratios, a threshold of 3.5
$\times$ below the S/N peak was chosen. Also the region around the CSPN was
excluded. For the shell, the same procedure was applied, but 20 $\times$
fainter.
}
\label{balmer_space}
\end{figure}

To search for intrinsic extinction in the nebula, the spatial
  distribution of the ratio $R$ was investigated. We used \ion{H}{$\alpha$}
  from the SAAO data, and \ion{H}{$\beta$} from the ESO data, as they are
  deeper in the outer shell (Fig. \ref{balmer_space}). The results show no
  correlation in the position and the line intensity. Even in the thin
  shell the same ratio was found.

Guerrero \& Manchado (\cite{gu_ma_99}) derive an extinction of $E_{\rm
  B-V}\,=\,0\fm12\pm0\fm03$. They note that their solution of the series does
not fulfill the theoretically expected values for such plasmas after
de-reddening. To obtain a reasonable signal to noise (S/N) ratio, they
integrate over large regions along the slit. The off-centered slit leads to
non-radial terms in the integrated line intensities. Taking into account the
offset of their slit, we find a fair agreement at the blue end.

\begin{figure}[!ht]
\centering
\includegraphics[bb=0 0 26.0cm 13.7cm, width=8.8cm]{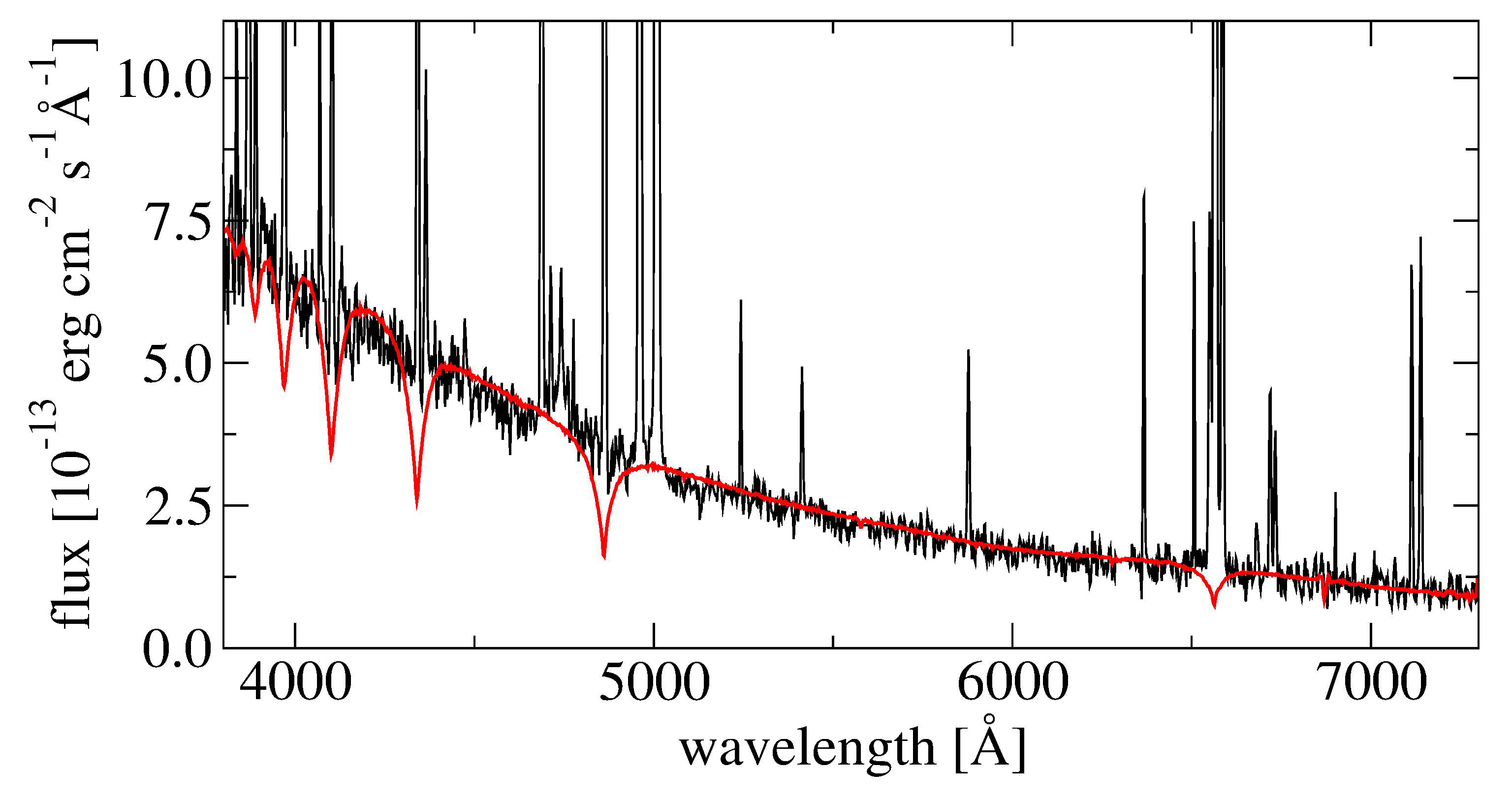}
\caption{The CSPN (thin black line), without removing the nebula emission
  lines, and the standard star EG~274 overlayed (smooth red line - scaled and
  the extinction $E_{\rm B-V} = 0\fm16$ applied).
}
\label{CSPN_vs_EG}
\end{figure}

A spectroscopic comparison of the hot white dwarf (WD) standard star
EG~274 and the CSPN leads to an independent test of the calibration. The CSPN
was completely covered by the ESO spectra and by three spectra (3, 4 and 9) in
the SAAO data set. The combined information from three different nights and
two different instruments reduces the calibration errors and excludes
systematic effects in the data reduction. The de-reddened SAAO spectrum,
overlayed with the WD spectrum of the standard star EG~274, shows the high
accuracy of the extinction value and of the calibration over the whole
wavelength range (Fig. \ref{CSPN_vs_EG}). Thus we were able to use this to
secure the photometry. The CSPN flux at $\lambda$5500\AA~ is
$2.3\,\times\,10^{-13}\,${erg cm}$^{-2}\,${s}$^{-1}\,${\AA}$^{-1}$ with an
{\sl rms} of 5\%. Using the zero point by Colina et al. (\cite{colinea96}), we
obtained values of V$\, =\,$16\fm82$\,\pm\,$0\fm09, and V$_0\,
=\,$16\fm32$\,\pm\,$0\fm10 for the CSPN.

\section{Distance determination}
\label{distance.sec}

There are several distance estimates of NGC 2438 in the literature, leading to
values between 0.8 and 4.3\ kpc.

Rauch et al. (\cite{rauch_99}) use the spectroscopy of the central star to
derive temperature and surface gravity. They use $V=18$\fm0, from Acker et
al. (\cite{esopn}), and a reddening in the range of $0\fm0~<~E_{\rm
  B-V}~<~0\fm34$. This leads to a large distance of 4.3~kpc and large
errors. Using the CSPN parameters and the extinction derived in this work (see
Sec. \ref{extinct_and_cspn}), we obtained a new spectroscopic distance of
$2.0^{+0.4}_{-0.35}$~kpc with their model.

For the open cluster M~46 at the same line of sight, a distance of 1.51 kpc
(Sharma et al. \cite{sharma_06}) is given from optical CCD photometry with
$E_{\rm B-V}=0\fm10\,\pm\,0\fm02$. Using infrared 2MASS data leads to an
extinction of $E_{\rm B-V}=0\fm13$ and a distance of 1.7$\pm$0.25\ kpc
(Majaess et al. \cite{majaess_07}). The abundance of [Fe/H]$=-0.03$ (Paunzen
et al. \cite{paunzen_10}) is very near to solar and the cluster is very
rich. The fits of the evolutionary tracks and thus the distances and
extinction values are reliable. Our extinction value places the PN behind the
cluster. This result agrees with Kiss et al. (\cite{kiss_08}), who already
rule out a cluster membership of the PN due to a large discrepancy in the
radial velocity.

Radio observations yield distance estimators based on a number of distance
scales with different assumptions and calibration. Using the high resolution
VLA radio map of Taylor \& Morris (\cite{vla}), we got a radio flux of
$74.9\,\pm\,1.6\,$mJy at a frequency of 1.5 GHz (20 cm). This is slightly
lower than the result of $80.3\,\pm\,3.2\,$mJy, given in the NVSS (Condon et
al. \cite{nvss}). The difference is due to the fact, that the background
source at 07$^{\rm h}$39$^{\rm m}$35\fs8s -14\degr28\arcmin04\arcsec was not
resolved by the survey. Using survey data from NVSS, PMN and F3R, the radio
spectral energy distribution (SED) can be fit by a pure free-free radiation (Vollmer et
al. \cite{specfind}). Thus using $I\,=\,I_0\,\nu^{-0.1}$ we derived a radio
flux of $S_{6\rm cm}\,=\,66\,$mJy. This corresponds well to the $S_{6\rm
  cm}\,=\,67\,$mJy, given by Zijlstra et al. (\cite{ZPB_89}) in their VLA
survey. Using the calibration of van de Steene \& Zijlstra (\cite{steene}) of
the radio continuum brightness temperature, we obtained a distance of
$1.8\,\pm\,0.3\,$kpc. The scale by Bensby \& Lundstr{\"o}m (\cite{BL_01})
results in $2.1\,\pm\,0.3\,$kpc. The errors mainly originate in the way we
determined the radius. We used the 20\% contour level of the \ion{H}{i} and
the [\ion{O}{iii}] image, averaged over the ellipticity. The errors are
estimates, varying inwards to the slightly smaller VLA image and outwards to
the 5\% level of the optical images. The statistical distance scale of
Schneider \& Buckley (\cite{schneiderbuckley}) results in a lower distance of
about 1.2~kpc. But such low distances tend to be excluded by the extinction
derived for the nebula and for the open cluster M~46.

Considering the individual spectroscopic distance and the distance of M~46 as
a foreground object, we adopt a distance of 1.9$\pm$0.2\ kpc for
NGC~2438. The error estimate makes use of the fact that Majaess et
  al. (\cite{majaess_07}) and Paunzen et al. (\cite{paunzen_10}) do not find
  intrinsic extinction variations in M~46, and thus NGC~2438 has to be beyond
  the cluster.  This is consistent with the nebula distance scale by
  van~de~Steene \& Zijlstra (\cite{steene}) and Bensby \& Lundstr{\"o}m
  (\cite{BL_01}).  Other statistical distance scales, giving results around
  1\,kpc, have to be excluded due to the values found for the open cluster
  M~46. The resulting dimensions of the nebula are summarized in
Tab. \ref{sizes}.

\begin{table}[t!]
\caption{Dimensions of the nebula, derived for a distance of 1.9\ kpc. }
\label{sizes}
\centering
\begin{tabular}{l c c c}
\hline\hline
 & measured & radius & dynamic age \\
 & diameter &\\
 & [\arcsec] & [$\times\,\,10^{17}$ cm] & [years]\\
\hline
   central wind cavity & ~~24 & ~~3.5 &  \\
   main nebula         & ~~76 & 10.8 & $12 \dots 17\,\times\,10^3$\\
   shell               & 144 & 20.5 &  \\
   first halo          & 189 & 27.~~ &  \\
   second halo         & 246 & 35.~~ &  \\
\hline
\end{tabular}
\tablefoot{The dynamical age agrees well with the 20~kyears derived as
  evolutionary age for the CSPN by Rauch et al. (\cite{rauch_99}).}
\end{table}

\section{Spectral analysis}

\subsection{General appearance}
\begin{figure}[!ht]
   \centering
   \includegraphics[bb=0 0 15.85cm 26.96cm, width=8.8cm]{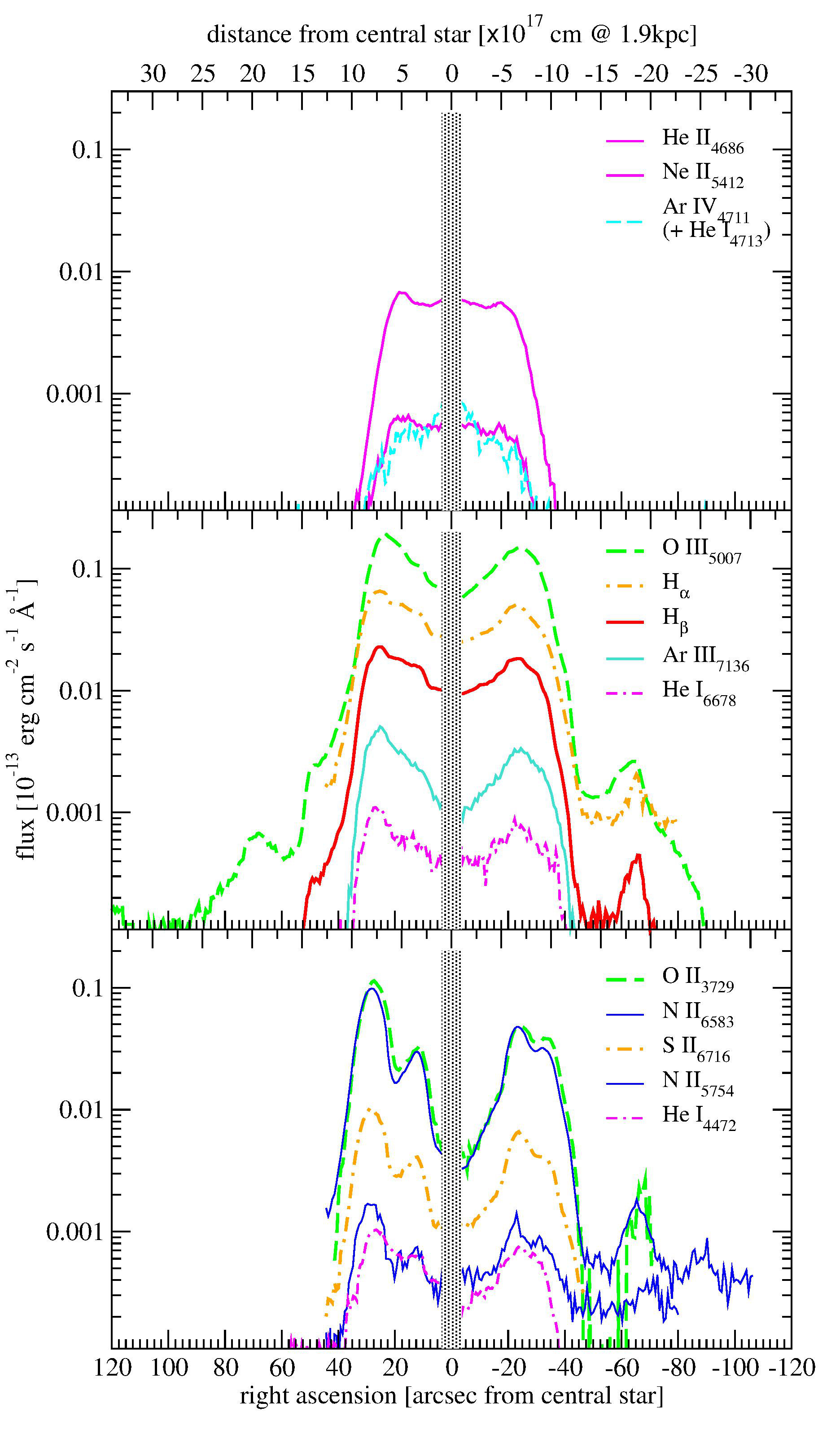}
      \caption{ Various lines of the nebula (logarithmic representation). The
        different spectral regions originate from different spectra (ESO \&
        SAAO) and thus vary in S/N ratio at similar physical intensities. The
        panels show the excitation groups (criteria see text).
        The high excitation species (upper panel) concentrate to the central region.
        The medium excitation species (middle panel) are smoothly following the
        total brightness and the radial behaviour of the VLA image throughout the nebula.
        The low excitation species (lower panel) are dominated by clumpy structures.
        The \ion{O}{ii} $\lambda$3729 is at the edge of the CCD and
          its flux calibration is very uncertain. It is added only for
          comparison of the profiles. }
      \label{all_lines.fig}
\end{figure}

The combined nebular spectra at the slit position through the CSPN are shown in
Fig. \ref{all_lines.fig}. The spectra obtained at the other positions were
compared to the combined spectra at the center, as a function of distance from
the CSPN. The variations between different lines are slight. Only the clumpy
low excitation species [\ion{N}{ii}] and [\ion{S}{ii}] vary up to a factor of
2 in their line ratios. This compares well with the results of the imaging
and the radio observations (see Fig. \ref{imaging}). The profiles show three
groups:
 \begin{itemize}
 \item The high excitation species (level $>$ 30 eV or ionization $>$ 60 eV) are concentrated to the center.
 \item The medium excitation species (level $>$ 10 eV or ionization $>$ 20 eV) are smoothly distributed over the whole
   nebula.
 \item The low excitation species (levels $<$ 5 eV and single ionized or neutral) are dominated by clumps.
 \end{itemize}
Only the triplet \ion{He}{i}$\lambda$4472\AA\ line and the singlet \ion{He}{i}$\lambda$6678\AA\ lines, due to the extreme long lifetime of the \ion{He}{i} $^3$S (19.82 eV) state of the singlet, faking a kind of ground state, behave not straight forward to the criteria given above.

\noindent Although \ion{He}{ii} and [\ion{O}{iii}] are coupled via the Bowen
fluorescence, the spatial profiles of these two elements differ. But within
one group, the profiles are identical.

\noindent These groups and the behavior of the low excitation species are
similar to those described by Ramos-Larios \& Phillips (\cite{NGC2371}) for
the PN  NGC~2371. It also resembles the \ion{H}{$\alpha$} : [\ion{N}{ii}]
surface brightness profiles of NGC~6369 (Ramos-Larios et
al. \cite{NGC6369}). The change of the excitation gives strong constraints for
the optical thickness in the models (see Sect. \ref{model.sec}).

\subsection{Density distribution and electron temperature}
\begin{figure}[th!]
   \centering
   \includegraphics[bb=0 0cm 18.7cm 16.2cm, width=8.8cm]{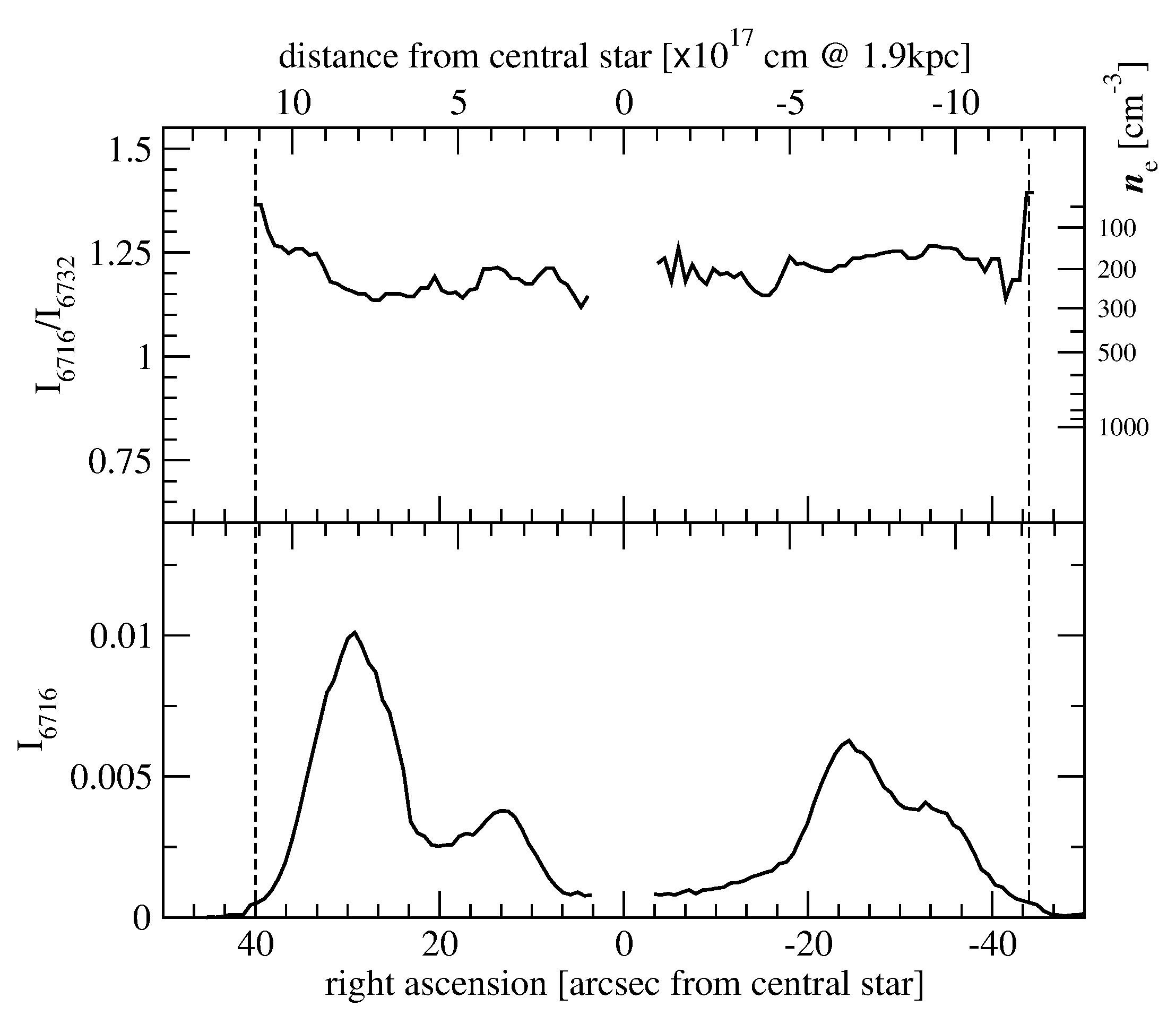}
\caption{The density derived from [\ion{S}{ii}], in E-W direction through the
  CSPN. The line ratio (upper panel) and the electron density is nearly
  constant. The density profile does not follow the strong surface brightness
  variations (lower panel).
      }
         \label{sii.fig}
\end{figure}

The diagnostic density and temperature diagrams (described in Osterbrock \&
Ferland (\cite{osterbrock}) and Proxauf et al. (\cite{diagnostic})) provide
the input and boundary conditions. Despite the strong variation in intensity,
the density derived with [\ion{S}{ii}] results in a flat distribution
(Fig. \ref{sii.fig}), with values of $200 \le n_{\rm e} \le
300$~cm$^{-3}$. This corresponds well to the value of 300~cm$^{-3}$,
calculated by Meatheringham et al. (\cite{vexp_88}), using the \ion{O}{ii}
$\lambda$3727\AA/$\lambda$3729\AA~ doublet and integrating over the whole
nebula. Both detections suffer from the fact that only low excitation species
were used.

\noindent Near the peak intensity of the main nebula, we could even detect
\ion{Cl}{iii} $\lambda$5517\AA/$\lambda$5537\AA. Throughout a large fraction
of the whole nebula, we found \ion{Ar}{iv}
$\lambda$4711\AA/$\lambda$4740\AA. Both doublets reflect the density of the
high excitation plasma. However, as shown in Stanghellini \& Kaler
(\cite{density_89}) and in Copetti \& Writzl (\cite{density_02}), those lines
are only suited for slightly higher densities. An upper limit of the density
of $300~ $to$ ~400$~cm$^{-3}$ is stated.

\begin{figure}[ht!]
   \centering
   \includegraphics[bb=0 0 25.3cm 17.4cm, width=8.8cm]{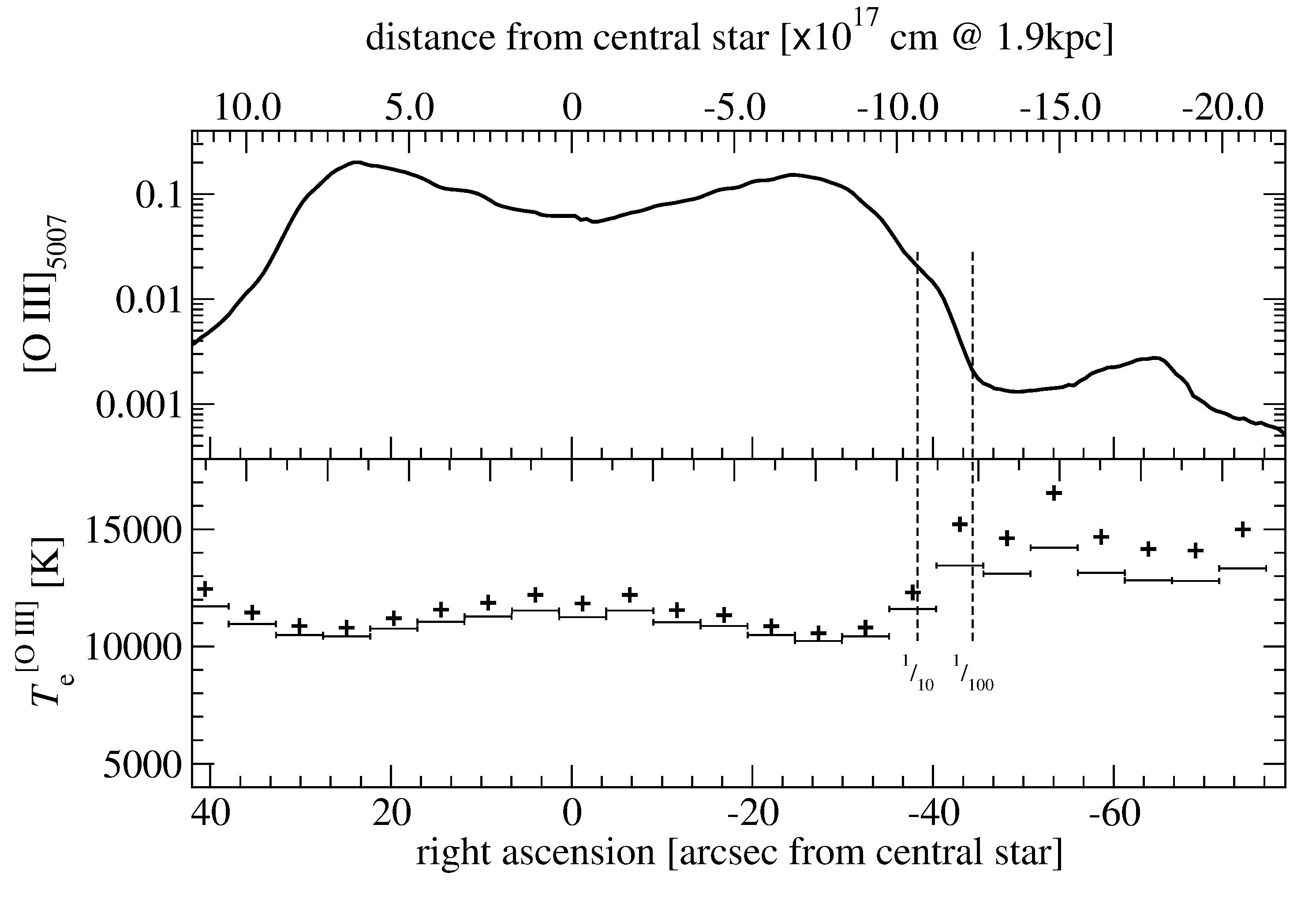}
      \caption{
   The intensity (upper panel) of the [\ion{O}{iii}] $\lambda$5007\AA, and the
   electron temperature using [\ion{O}{iii}]
   ($\lambda$4958\AA+$\lambda$5007\AA)/$\lambda$4363\AA~ ratios along the slit
   (lower panel). Logarithmic scale is used to show the shell and the even
   further extended halo. The flux scale is the same as in
   Fig. \ref{all_lines.fig}. The plus signs indicate the solution by
   exponential formula of Osterbrock \& Ferland (\cite{osterbrock}). The bars
   give the solution using the new calibration of Proxauf et
   al. (\cite{diagnostic}).}
   \label{oiii_temp.fig}
\end{figure}

The [\ion{O}{iii}] ($\lambda$4958\AA+$\lambda$5007\AA)/$\lambda$4363\AA~ratios
along the slit result in an electron temperature of 10\,000 to 11\,500~K in
the main nebula. A rise in the shell is suggested (see
Fig. \ref{oiii_temp.fig}). The transition to the shell is marked by dashed
lines with $^1/_{10}$ and $^1/_{100}$ of the peak intensity. The result
reaches up to 13\,000~K, using the new calibration of Proxauf et
al. (\cite{diagnostic}). Because of the weakness of the [\ion{O}{iii}]
$\lambda4363$\AA, which is about 100 times fainter compared to the
[\ion{O}{iii}] $\lambda5007$\AA~ emission, this may be an overestimation.
There is no indication for a significant decline in temperature, as it would be
expected for a recombining halo. The presence of the $\lambda4363$ line ($> 5eV$)
confirms a high temperature.

\noindent Only in the main nebula we are able to derive the [\ion{N}{ii}]
($\lambda$6548\AA+$\lambda$6583\AA)/$\lambda$5755\AA~ line ratios. The results
are very close to that of [\ion{O}{iii}] (see Tab. \ref{model_input.tab}). This
implies that we  see the same material of the nebula in two different
excitation states. A third temperature diagnostic is given by Keenan et
al. (\cite{T_ar_88}), using \ion{Ar}{iii}
($\lambda$7135\AA+$\lambda$7751\AA)/$\lambda$5192\AA. Unfortunately, our
spectral resolution is insufficient to detect \ion{Ar}{iii} $\lambda$5192\AA~
in the blend with the \ion{N}{i} $\lambda$5198\AA+$\lambda$5200\AA~ lines.

\subsection{Searching for shock signatures}
\label{shock.sec}

As the radiative transfer hydrodynamic (RTH) models predict shocks during certain stages of the evolution
(Corradi et al. \cite{corradi_00}, Sch\"{o}nberner \& Steffen
\cite{schoenberner_02}, Perinotto et al. \cite{perinotto_04}), we searched
carefully for shock signatures, to exclude possible regions where the
photoionization model does not work.
\begin{figure}[!bht]
   \includegraphics[bb=0 0 27.2cm 17.4cm, width=8.8cm]{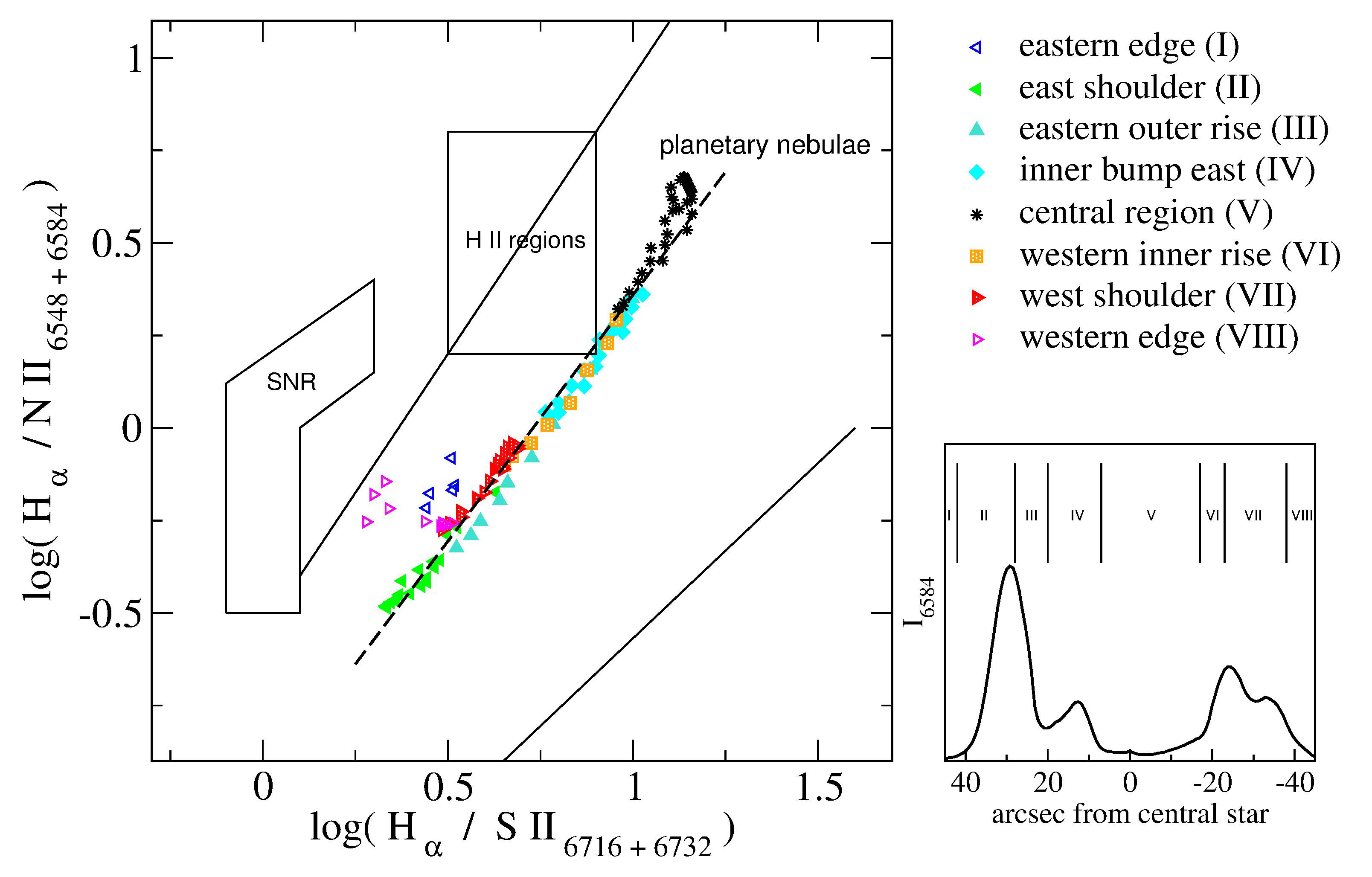}
      \caption{The diagnostic diagram of the separation of shocked nebulae
        from photoionized PNe (after Magrini et al. \cite{magrini_03}). The
        points follow the sequence of the ionization class towards the
        center. The dashed gives a fit through the good data points (filled symbols).
        No dependency on rise or fall of the intensity can be found
        (small panel).
      }
         \label{ha_sii_nii.fig}
\end{figure}

Originally shown by Garc{\'{i}}a Lario et al. (\cite{garcia_lario}), and later
investigated more in detail by Magrini et al. (\cite{magrini_03}), the ratios
of $\log($\ion{H}{$\alpha$}/[\ion{N}{ii}]$)$
vs. $\log($\ion{H}{$\alpha$}/[\ion{S}{ii}]$)$ can be used to separate
photoionized PN plasma from shocked gas, as seen, e.g., in supernova remnants
(SNRs). Schmeja \& Kimeswenger (\cite{schmeja}) have shown that, for
photoionized gas of symbiotic Miras, the position in the diagram is an
indicator for the level of excitation as well.
The spectra of the main nebula of NGC~2438 show the excitation
tendency from inside to outside well (Fig. \ref{ha_sii_nii.fig}).There is no
signature of a deviation from expected line ratios. This is an indication that
photoionization is the dominating mechanism for the excitation of the
material.  The fit line, used for the further analysis, has an inclination of 1.33 and a
constant of 0.73 along the abscissa. A few data points at the outermost edge suffer from weak sulphur
lines. At least one order of magnitude deeper spectra would be required to
extend this kind of analysis towards the shell.
\begin{figure}[th!]
   \centering
   \includegraphics[bb=0 0 8.8cm 9.03cm, width=6.5cm]{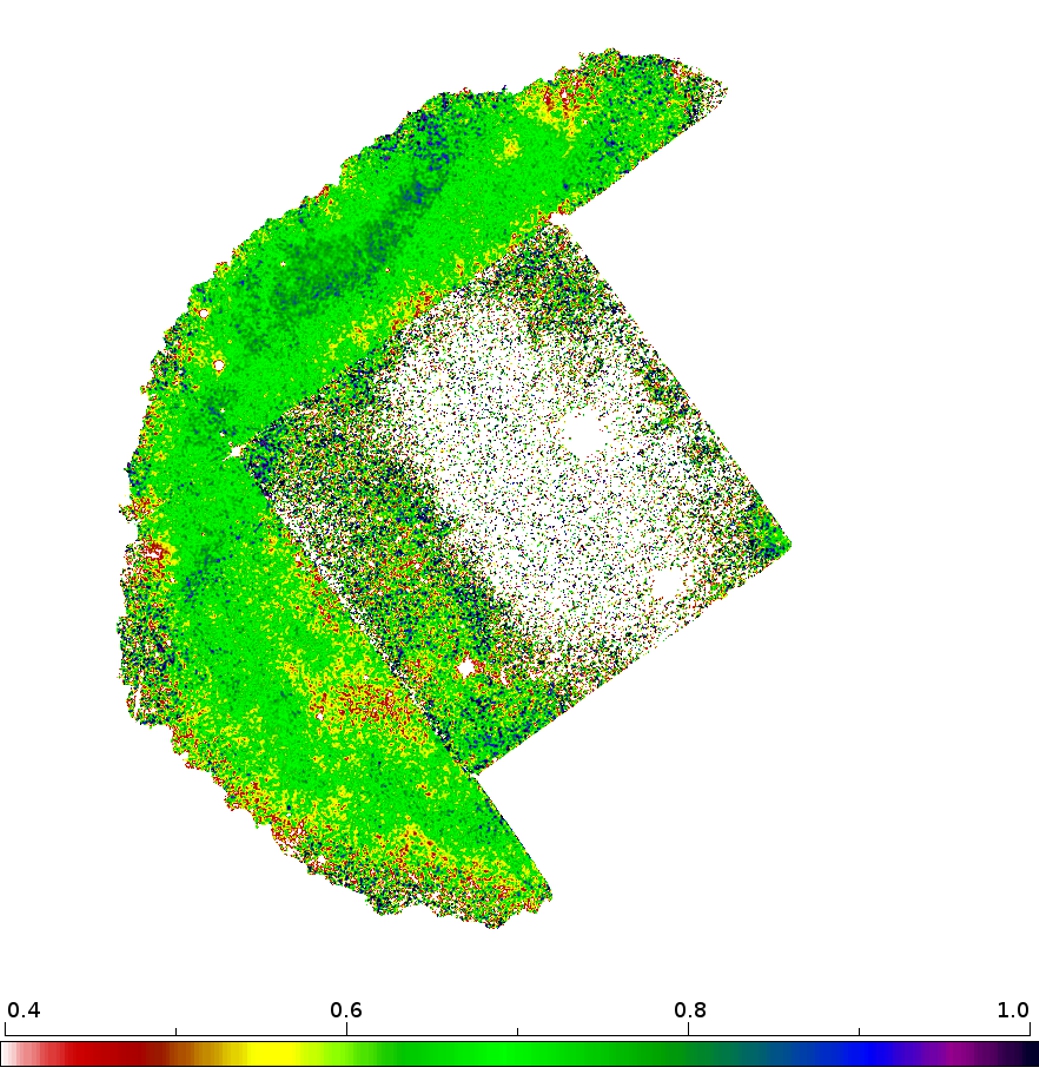}
\caption{The $\log({\rm H}\alpha/{\rm [\ion{N}{ii}]}) - \log({\rm H}\alpha/{\rm [\ion{N}{ii}]})$ image obtained from the HST frames. It is based on the same analysis as in Fig.\ \ref{ha_sii_nii.fig}. The expected value
  should be the same as the incline (0.7-0.8) of the trendline through the data points in Fig.\ \ref{ha_sii_nii.fig}.
        }
\label{hst_magrini.fig}
\end{figure}
\noindent Although the HST images do not cover the whole nebula and are very
noisy (especially in [\ion{S}{ii}] in the PC camera part), they were used to
obtain a similar analysis. The linear fit in
Fig. \ref{ha_sii_nii.fig} has a slope of 1.33. The HST Filter F658N
covers only the [\ion{N}{ii}] $\lambda$6584\AA~line and the line ratio of
[\ion{N}{ii}] $\lambda$6584\AA~ is known to be 1:3, so the factor goes to
unity. As both [\ion{S}{ii}] lines are covered by the HST filter, the
abscissa stays untouched. The subtraction $\log($\ion{H}{$\alpha$}/[\ion{N}{ii}]$)-
\log($\ion{H}{$\alpha$}/[\ion{S}{ii}]$)$ thus should result in the constant
along the abscissa, if the whole nebula follows the tendency given in the region covered
by the spectra. Fig.\,\ref{hst_magrini.fig} clearly shows this behavior. As
outer boundary the level in the \ion{H}{i} image at 2.5\% of the peak
intensity was chosen.

\begin{figure}[!ht]
   \centering
   \includegraphics[bb=0 0 8.8cm 9.03cm, width=6.5cm]{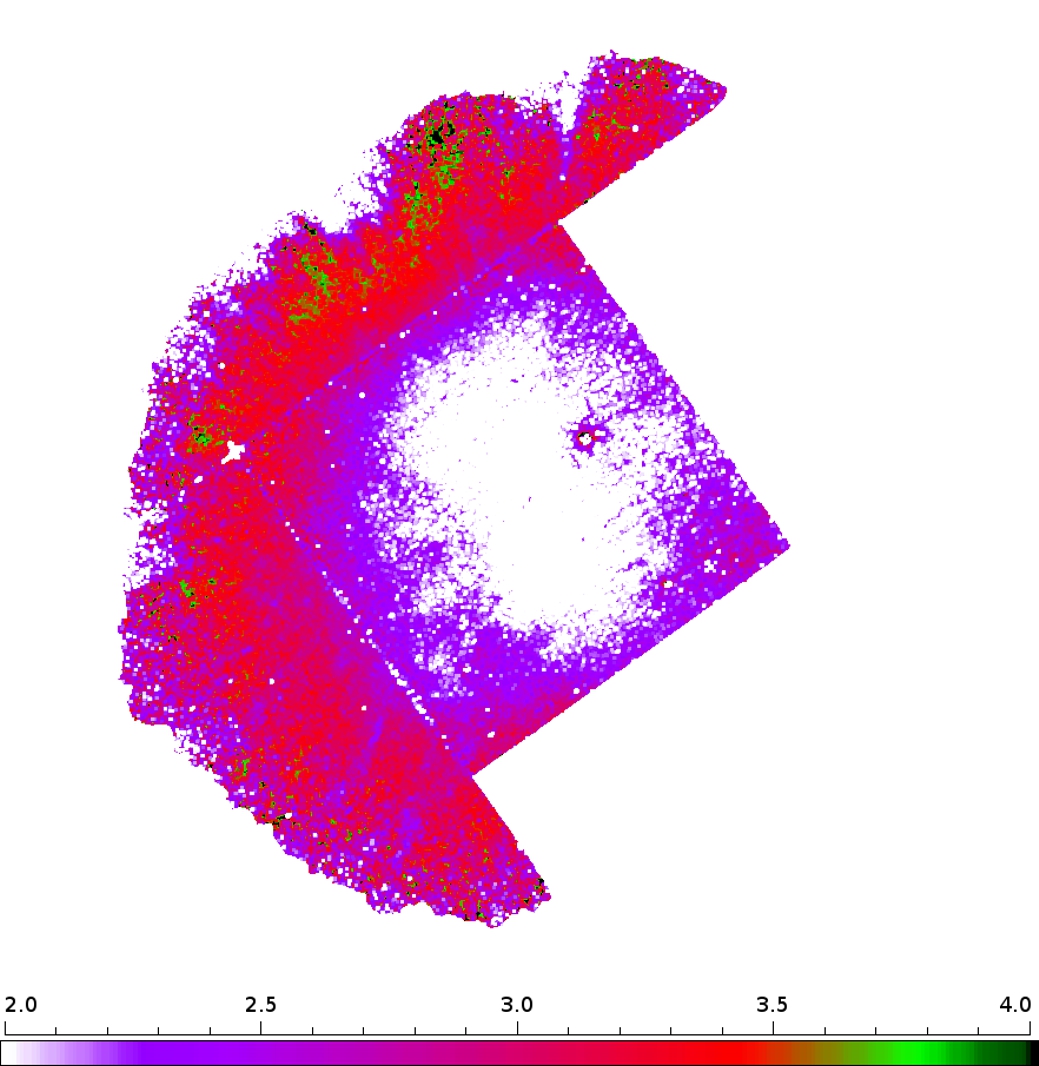}
\caption{The [\ion{O}{iii}]/\ion{H}{$\alpha$} image from the HST data. No
  sudden changes in the line ratio can be found.
      }
\label{hst_guerrero}
\end{figure}

Recently, Guerrero et al. (\cite{guerrero_13}) suggested to use the
[\ion{O}{iii}]/\ion{H}{$\alpha$} ratio to unveil shocks in regions of higher
temperature. In those regions, nitrogen and sulfur are at least double
ionized, and thus, no strong [\ion{N}{ii}] and [\ion{S}{ii}] lines are
expected. They show that shocks should indicate sudden changes of the line
ratio. They use similar HST archive data, but excluded our target due to
possible contamination in the F656N filter with nitrogen lines. As discussed
in Sec~\ref{imaging.sec}, in this case the images taken with F658N can be used
in order to correct for contamination. Fig.\,\ref{hst_guerrero} shows the
result for our target. The line ratio is fairly constant: no shock signatures
could be found in this target in the region covered by the images.

\section{Model and discussion}
\label{model.sec}
All calculations were performed with version C10 of CLOUDY (Ferland et
al. \cite{cloudy}). After the release of version C13 (Ferland et
al. \cite{cloudy13}), the resulting models were recalculated with the new
version. No differences were identified between all results. The density
profile was parameterized in 12 shells of $5\,\times\,10^{16}$\,cm. The nomenclature used here is the same as mentioned in Sect. 1.

%\relax
\subsection{The main nebula}

In the main nebula, the strong variation of the [\ion{S}{ii}] flux (see
Fig. \ref{sii.fig}) and, on the other hand, the constant density given by the
diagnostic diagrams require a change of the filling factor in the middle of
the main nebula. This has been parameterized with two free parameters, giving the
filling factors inside and outside a critical distance. The latter was
identified by changes in the direct images.
This is similar to the observed changes in clump frequencies in the Helix
Nebula, which was recently discovered to be also a multiple-shell planetary
nebula (MSPN) (Zhang et
al. \cite{zhang12}). The optical images of the Helix Nebula do not cover all
radii (O'Dell et al. \cite{helix_05}) and show a difference in number density
at a line about 40\% of the radius for the large knots. Matsuura et
al. (\cite{helix_09}) describe a low frequency of isolated H$_2$ clumps in the
inner half of the nebula and a strong increase to a few hundred isolated knots
per square arcminute in a region they call the inner ring (out to about 70\%
of the radius). After that region, a dense, cloudy structure abruptly starts,
with almost no individual, isolated knots with tails, but with a nearly
complete surface coverage. According to a comparison with a model originally
computed for NGC 2436 (Vicini et al. \cite{vicini_99}) by Speck et
al. (\cite{ring_03a}), the Helix Nebula should be just on the edge to destroy
the cold molecular knots, with it's age of 19\,000$^{+10\,000}_{-8\,000}$
years very similar to our target. In the Ring Nebula, Speck et
al. (\cite{ring_03a}) and O'Dell et al. (\cite{ring_03b}) find similar overall
behavior. But the Ring Nebula is assumed to be much younger
(5\,000$^{+2\,500}_{-1\,700}$ years) and may still be evolving, and thus by a
ploughing effect colliding smaller knots may combine to bigger ones or even more knots and clumps may be formed. Speck et al.
(\cite{ring_03a}) also show, in their narrow band
imaging, that the transition to the region with a high number density of no
longer individual H$_2$ clouds is coincident with a sudden drop in the
\ion{He}{ii}/[\ion{O}{iii}] ratio. In our investigation, the sudden drop of
the \ion{He}{ii} $\lambda$4686\AA~ demands such a change as well. The position
of the change of the filling factor is thus fixed at $6\,\times\,10^{17}$\,cm.

\begin{figure}[!ht]
   \centering
   \includegraphics[bb=0 0 26.6cm 20.7cm, width=8.8cm]{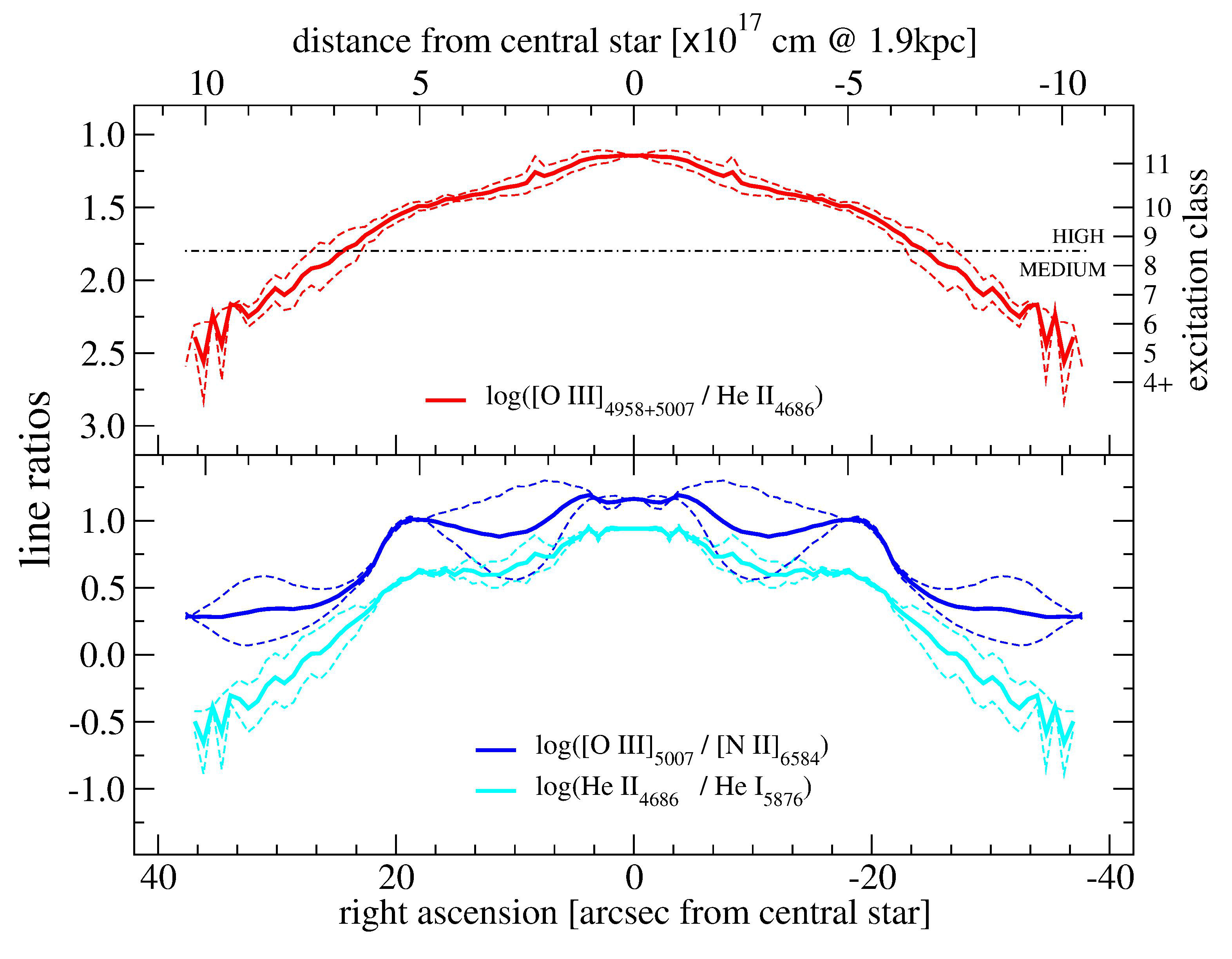}
      \caption{The excitation and ionization sensitive line ratios. The dashed lines
      are the spectra along the slits through the center in original
      orientation (E-W) and the same spectrum reversed in direction to see the asymmetries. The thick lines are the average of both
      directions. The excitation class and the level of separation of medium
      to high excitation is used according to the calibration by Reid \&
      Parker (\cite{Reid_2010}).}
         \label{excitation.fig}
\end{figure}

An accurate model of the CSPN is essential for an appropriate modeling of the
whole nebula. As shown by Armsdorfer et al. (\cite{armsdorfer_02}), the
influence of the chosen CSPN model is crucial, in particular for the strength
of the helium lines. Thus we used state-of-the-art H-Ni NLTE CSPN models
provided by the T{\"u}bingen group (Rauch \cite{rauch_03}, Ringat
\cite{ringat_12}). The stellar parameters given by Rauch et
al. (\cite{rauch_99}) were used. The temperature was varied within one sigma
of their original result.  The spatial distribution of the line ratios,
  used for excitation and ionization as discussed in Miko{\l}ajewska et
  al. (\cite{CVs}), Phillips (\cite{Phillips_2004}) and Reid \& Parker
  (\cite{Reid_2010}), were used for an initial guess of the fitting procedure
  (see Fig.~\ref{excitation.fig}). The often used line ratios including
  [\ion{O}{ii}] were not used here, as this [\ion{O}{ii}] line is sensitive to
  density variations. Additionally, the line is just at the edge of the CCD
  chip and therefore less reliable.

\begin{figure*}[!ht]
   \centering{\includegraphics[bb=0 0 26.96cm 16.93cm, width=18cm]{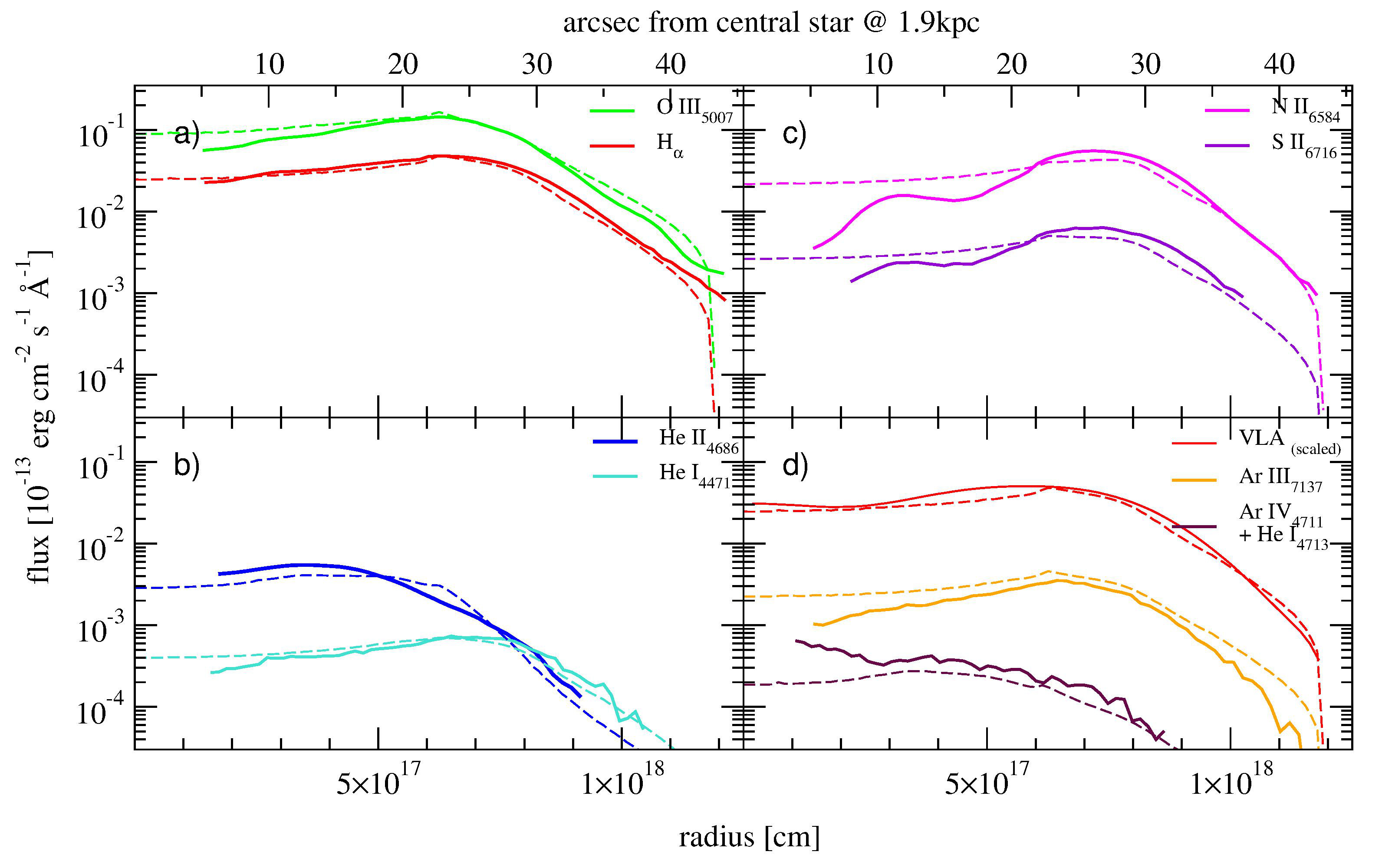}}
      \caption{The complete CLOUDY model of the main nebula. The dashed lines give the observational data, while the full lines are giving the model. The spectral lines
        in the left panels (a) and (b) are used to fit the free parameters of
        the density law, the filling factor and the temperature of the
        CSPN. The right panels (c) and (d) give the results derived for other
        lines from the best fit model. The VLA intensities were arbitrary scaled up to the same level as the peak in H$\alpha$.}
         \label{model.fig}
\end{figure*}

\begin{table}[!ht]
\caption{ The parameters of the target and of the best fit model of the main
  nebula (14\arcsec-38\arcsec)} at three projected radii.
\label{model_input.tab}
\centering
\begin{tabular}{llcc cc cc}
\hline\hline
\multicolumn{4}{l}{Distance:\,\,$D$}&\multicolumn{2}{l}{1.9 $\pm$ 0.2\,$[$kpc$]$}\\
\multicolumn{4}{l}{$E_{\rm B-V}$}&\multicolumn{2}{l}{0\fm16 $\pm$ 0\fm01}\\
\multicolumn{4}{l}{Diameter:\,\,$d_{\rm main nebula}$} & \multicolumn{2}{l}{0.65\,$[$pc$]$}\\
\multicolumn{4}{l}{Mass:\,\,$M_{\rm main nebula}$} & \multicolumn{2}{l}{0.45\,M$_\odot$}   \\
\\
\multicolumn{4}{l}{CSPN:  Rauch et al. (\cite{rauch_99})}& &\\
\multicolumn{4}{l}{$T_{\rm CSPN}$} & \multicolumn{2}{l}{114 $\pm$ 10\,$[$kK$]$}   \\
\multicolumn{4}{l}{$\log(g[\rm cgs])$} &
\multicolumn{2}{l}{$6.62\pm0.22$}\\
\multicolumn{4}{l}{$L_{\rm CSPN}$}&\multicolumn{2}{l}{$570\,L_\odot$}\\
\multicolumn{4}{l}{$M_{\rm CSPN}$}&\multicolumn{2}{l}{$0.56\pm0.01\,\,M_\odot$}\\
\multicolumn{4}{l}{$\log(n_{\rm He}/n_{\rm H})$}&\multicolumn{2}{l}{$-0.56\pm0.27$}\\   \\
\multicolumn{4}{l}{CSPN:  model}\\
\multicolumn{4}{l}{$T_{\rm CSPN}$} & \multicolumn{2}{l}{120 $\pm$ 2\,$[$kK$]$} \\
\multicolumn{4}{l}{$L_{\rm CSPN}$} & \multicolumn{2}{l}{$550 \pm 15\,L_\odot$}   \\
\hline
\multicolumn{2}{l}{projected} & \multicolumn{2}{c}{$3.5\times10^{17}$cm} & \multicolumn{2}{c}{$6.5\times10^{17}$cm} &
\multicolumn{2}{c}{$9.0\times10^{17}$cm} \\
\multicolumn{2}{l}{\phantom{xxi}radius} & \multicolumn{2}{c}{$\equiv$13\arcsec} & \multicolumn{2}{c}{$\equiv$23\arcsec} &
\multicolumn{2}{c}{$\equiv$32\arcsec} \\
\multicolumn{2}{l}{line ($\lambda\,[$\AA$])$} & obs. & model & obs. & model & obs. & model \\
\hline                        % inserts single horizontal line
\multicolumn{2}{l}{$n_{\rm H}$}              &         & 160     &         & 295     &         & 210 \\
\multicolumn{2}{l}{$n_{\rm e}$([\ion{S}{ii}])} & 180     & 200     & 350     & 300     & 200     & 180  \\
\multicolumn{2}{l}{$\log(n_{\rm He}/n_{\rm H})$}             &         & -0.56    &         & -1.0    &         & -1.0 \\
\multicolumn{2}{l}{$T_e$(phys.)}             &         & 11.7    &         & 10.2    &         & 10.6 \\
\multicolumn{2}{l}{$T_e$([\ion{O}{iii}])}
   $\!\!\!\left.{\parbox[c]{0.25cm}{
\tiny a)
\tiny b)}}\right\}$
   &
   $\!\!\left\{{\parbox[c]{0.40cm}{
11.3
10.8}}\right.$
   & 10.8    &
   $\!\!\left\{{\parbox[c]{0.40cm}{
10.9
10.5}}\right.$
    & 10.0    &
    $\!\!\left\{{\parbox[c]{0.40cm}{
11.6
11.1}}\right.$
    & 10.4  \\
   & & \multicolumn{6}{l}{\tiny a) calibration by Osterbrock \& Ferland (\cite{osterbrock})} \\
   & & \multicolumn{6}{l}{\tiny b) calibration by Proxauf et al. (\cite{diagnostic})} \\
\multicolumn{2}{l}{$T_e$([\ion{N}{ii}])}    & 10.5    & 10.1    & 10.0    & 10.0    & 13.3    & 10.3  \\
\multicolumn{2}{l}{filling factor}            &         & 0.10    &         & 0.38    &         & 0.38 \\
\hline
\end{tabular}
\tablefoot{The projected radii  correspond to the peaks of \ion{He}{ii},
  \ion{H}{$\rm \beta$} and the outer edge ( \ion{H}{$\alpha$} $<$ 25\% of
  peak).
  %The intensities are given relative to the model intensity of
  %\ion{H}{$\rm \beta$} at $6.5\,\,10^{17}$cm (=100).
  The densities are given
  in $[$cm$^{-3}]$.}
\end{table}

\noindent Only the spatial profiles of [\ion{O}{iii}] $\lambda$5007\AA,
\ion{H}{$\alpha$}, \ion{He}{i} $\lambda$4471\AA~ and \ion{He}{ii}
$\lambda$4686\AA~ were used to fit the free parameters of the nebula.
For helium, the $\lambda$4471\AA~ line was used instead of the brighter $\lambda$5876\AA~ line,
because it was covered by the significantly deeper ESO spectra.
The fitting of the stellar temperature is completely dominated by the \ion{He}{i}
to \ion{He}{ii} line ratio. The \ion{He}{ii} to [\ion{O}{iii}] ratio in the
inner region can be described only with the CSPN model by Rauch et
al. (\cite{rauch_99}). This suggests, that the inner region is dominated by
the products of the stellar winds from the current photosphere of the
CSPN. All other abundances were taken from the intrinsic set for PNe in CLOUDY
by Aller \& Czyzak (\cite{abundance_83}) and Khromov
(\cite{abundance_89}). There were not enough lines to do detailed, spatially
resolved abundances for other elements in this fitting procedure as free
parameters. The best fit model is summarized in Tab.\,\ref{model_input.tab}.

\noindent After fitting the whole set of free parameters, all other detected
lines in the spectra are calculated (Fig.~\ref{model.fig},
Tab.~\ref{model_list.tab}). There is no need for any iteration or feedback on
the model. Even within the errors, no major deviations from the abundances
seem to be required.

\noindent The simple model of the VLA data (see
Fig.~\ref{model.fig}) was produced by using the free electrons of hydrogen and
helium and assuming only free-free radiation. The radio flux density was scaled to the same level of peak as the
\ion{H}{$\alpha$} line, to fit into the scale of the plot. Additionally it is
smeared with a gaussian, representing the beam size given by Taylor \& Morris
(\cite{vla}). Also shown in Fig.~\ref{model.fig}, the feature around
$\lambda$4712\AA~ can not be described by [\ion{Ar}{iv}]+\ion{He}{i}. At the
inner region of the hot wind bubble, a high excitation species additionally
seems to contribute to the wind. This region was not modeled in this setup. A
possible candidate might be \ion{Ne}{iv}, although the contribution calculated
by CLOUDY lies a factor of 40 below the other two lines in the main nebula.

\noindent The resulting integrated mass of the main nebula model is
0.45~M$_\odot$. This is a high, but still feasible mass for a PN.

\begin{table}[!ht]
\caption{The parameters of the best fit model of the shell (38\arcsec-72\arcsec).}
\label{shell_input.tab}
\centering
\begin{tabular}{llcc cc cc}
\hline\hline
\multicolumn{2}{l}{Parameter} &
\multicolumn{2}{c}{\phantom{$3.5\times10^{17}$cm}} & \multicolumn{2}{c}{\phantom{$6.5\times10^{17}$cm}} &
\multicolumn{2}{c}{\phantom{$9.0\times10^{17}$cm}} \\
\hline
\multicolumn{4}{l}{Diameter:\,\,$d_{\rm shell}$\phantom{XXXXXX}} & \multicolumn{2}{l}{1.25\,$[$pc$]$}\\
\multicolumn{4}{l}{Mass:\,\,$M_{\rm shell}$} & \multicolumn{2}{l}{0.5\,\dots\,0.8\,M$_\odot$}   \\
\multicolumn{4}{l}{Filling factor} & \multicolumn{2}{l}{0.5\,\dots\,1.0}   \\
\multicolumn{4}{l}{$n_{\rm H}$} & \multicolumn{2}{l}{7\,\dots\,30\,cm$^{-3}$} \\
\multicolumn{4}{l}{$n_{\rm H II}/n_{\rm H}$} & \multicolumn{2}{l}{$\ge\,0.97$}   \\
\multicolumn{4}{l}{$T_e$([\ion{O}{iii}]) observed} & \multicolumn{2}{l}{12.5\,\dots\,14.2\,[kK]}   \\
\multicolumn{4}{l}{$T_e$ model} & \multicolumn{2}{l}{13.2\,[kK] (fixed)}   \\
\hline
\end{tabular}
\tablefoot{The linear diameters correspond to the derived distance of 1.9\,kpc.}
\end{table}

\subsection{The shell}

The sensitivity of the VLA observations, as the electron density in the shell
is about two orders of magnitude below that of the main nebula, was unfortunately insufficient to
obtain a direct measurement, and diagnostic diagrams using forbidden lines
saturate below $n_{\rm e} \approx 50 \dots 100$~cm$^{-3}$. So we had to derive
model calculations. Only the lines of \ion{H}{$\alpha$}, \ion{H}{$\beta$},
[\ion{O}{iii}] $\lambda$5007\AA~ and [\ion{N}{ii}]
$\lambda$6548\AA+$\lambda$6584\AA~ are sufficiently bright for
the profile analysis. [\ion{O}{iii}] $\lambda$4363\AA~ was detected close to
the noise level. Thus the average value of the electron temperature of
13\,200K (see Fig.\,\ref{oiii_temp.fig}) between 42\arcsec and 60\arcsec was
used, instead of a free input parameter of the fit.

\begin{figure}[!ht]
   \centering
   \includegraphics[bb=0 0 27.4cm 20.3cm, width=8.8cm]{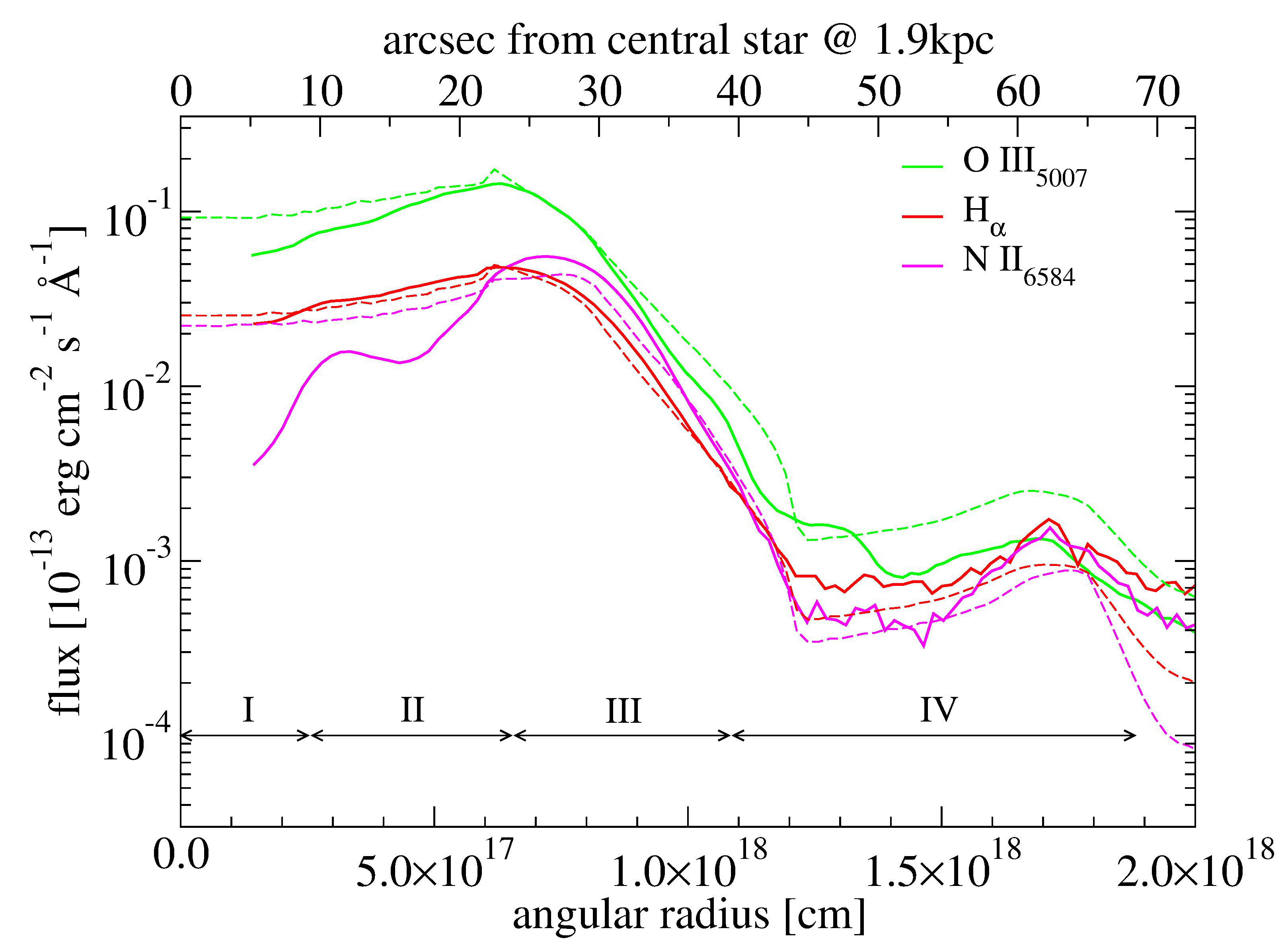}
      \caption{The CLOUDY model of the shell (dashed lines = observations, full lines = model).  As just a few lines could be
        used, it has the density and the filling factor as single free
        parameters (see text). The inner regions (I - III) are not fitted in
        this model, but taken from the main nebula model.}
         \label{model_shell.fig}
\end{figure}

The CLOUDY model of the shell (Fig. \ref{model_shell.fig}) is using the
transmitted radiation of the main nebula model. It requires a density of
$n_{\rm H} \approx 7 \dots 30$~cm$^{-3}$ (see Tab. \ref{shell_input.tab}). The
\ion{H}{$\alpha$} / [\ion{N}{ii}] line ratio is slightly above unity. This
ratio can be achieved only by a filling factor higher than the one in the main
nebula for this plasma. The calculations provide nearly the same results
for the three lines in the whole range:
0.5~$\le$~filling~factor~$\le$~1.0. Thus we are unable to give a good
conclusion about this parameter.

\subsection{Comparison to the RTH models}

The radiative transfer hydrodynamic (RTH) models calculated for NGC\,2438 by Corradi et al. (\cite{corradi_00}),
Sch\"{o}nberner \& Steffen (\cite{schoenberner_02}) and Perinotto et
al. (\cite{perinotto_04}) lead to different results in some aspects.

\noindent The results of a 1D RTH calculation by Corradi et
al. (\cite{corradi_00}) and Sch\"{o}nberner \& Steffen
(\cite{schoenberner_02}) show mainly recombining gas in the shell. The electron
number density is one order of magnitude below that of hydrogen. In the RTH
models, a temperature of  2\,000 to 3\,000~K is indicated. So we tested
various setups with fixed electron temperatures down to 2\,000 to 3\,000~K,
but they did not result in a fit of the ratio of the observed lines at a
reasonable density. Additionally, there is no detection of the [\ion{O}{i}]
lines (although those lines are affected by substraction of the Earth's
aurora). After applying estimates on the strengths and variation of the telluric
lines by the methods described in Noll et al. (\cite{noll_12}), we are quite
  confident about this result. The ratio predicted by the RTH models is
  [\ion{O}{i}]~$\lambda~6300$\AA~:~\ion{H}{$\beta$}~=~6~:~1, but we could not
  confirm this result.

\noindent The resulting integrated mass of the main nebula model is 0.45~M$_\odot$ in
our calculations. It is about a factor of two above the mass calculated by
Corradi et al. (\cite{corradi_00}), integrating the RTH model at their adopted
distance of 1.0~kpc. At the recalculated distance of 1.9 kpc, their model
gives a  mass of 1.7~M$_\odot$.

\noindent The resulting integrated mass of the shell is about
0.5~to~0.8~M$_\odot$ in our determinations. The RTH models of Okorokov et
  al. (\cite{okorokov85}), Corradi et al. (\cite{corradi_00}) and Perinotto et
  al. (\cite{perinotto_04}) assume that the shell is AGB wind
  material. Calculations with such an AGB wind with a terminal velocity of
  10~km~s$^{-1}$ and a typical mass-loss rate of
  $2.5\,\times\,10^{-5}$~M$_\odot$~yr$^{-1}$ result in a mass of 0.8~M$_\odot$
  in the same volume, consistent with our results. The different density used
  in the RTH models of Corradi et al. (\cite{corradi_00}) and Perinotto et
  al. (\cite{perinotto_04}) leads to a very high mass estimate of
  3.5~M$_\odot$ at a distance of 1.9 kpc. This would require a very high mass
  loss of $\gtrsim\,1.1\,\times\,10^{-4}$~M$_\odot$~yr$^{-1}$ during the late
  AGB phase. At the smaller distance of 1 kpc, their calculations result in an
  integrated mass of about 0.5~M$_\odot$. But such small distances are
  excluded (see sect. \ref{distance.sec}).

The reason for the differences in the results may be found in the change of
the input parameters (e.g., distance or luminosity). The RTH models used older
input parameters in their calculations. Some of these parameters have changed
due to new measurements and calculations. An update of the RTH models to
include the new measurements would be of interest to test whether the results
would converge. The evidence for clumping may also complicate the
interpretation of the 1d RTH models.

\section{Conclusions}

The observations of NGC 2438 allow us to derive an individual distance of
$1.9\pm0.2\,$kpc and a foreground extinction of $E_{\rm
  B-V}=0\fm16\pm0\fm01$. We confirm its non-membership of the open cluster
M~46, in the foreground. The large discrepancy of the nebula luminosity and
the CSPN luminosity (Rauch et al. \cite{rauch_99}) is completely resolved with
these values.

The model of the main nebula indicates that the old MSPN is matter
bounded. The filling factor in the inner region is lower than that in the
outer part of the main nebula. This is very similar to the observational results of
the spatially resolved, younger Ring and Helix nebulae. Although using only
four lines in the parameter fitting, the photoionization model shows a nearly
perfect representation of all observed lines. The analysis of the shock
sensitive tracers indicates that shocks do not contribute to the excitation
of low ionized atoms like [\ion{N}{ii}] and [\ion{S}{ii}]. This old nebula is
dominated by photoionization.

The surface brightness distribution of a few bright lines by the RTH models
(Corradi et al. \cite{corradi_00}; Perinotto et al. \cite{perinotto_04}) lead
to a fair representation of the whole nebula. However, the excitation and
temperature throughout the nebula, and beyond, needs the handling of small scale
clumps to get a self consistent model. A combination of the sophisticated
hydrodynamical calculations in the RTH models, including effects of turbulence
and clump formation with photoionization, would be required for a complete
view.

The determined temperature of the shell and the missing bright [\ion{O}{i}]
lines lead to the conclusion that the shell of NGC 2438 is fully ionized.
Even the [\ion{O}{iii}] lines indicate a photoionized state. Final
confirmation by observations of shock sensitive lines (e.g. [\ion{S}{ii}]
$\lambda$6716\AA/$\lambda$6732\AA), to verify possible other excitations, is
still missing. More detailed investigations to derive electron temperatures
from lower ionized species (e.g.,
[\ion{N}{ii}]($\lambda$6548\AA+$\lambda$6584\AA)/[\ion{N}{ii}]$\lambda$5755\AA~
or
[\ion{O}{ii}]($\lambda$3727\AA+$\lambda$3729\AA)/[\ion{O}{ii}]($\lambda$7320\AA+$\lambda$7330\AA)),
and from elements not influenced by depletion in dust grains, like
[\ion{Ar}{iii}] $\lambda$5192\AA/($\lambda$7135\AA+$\lambda$7751\AA), are highly
desired for the main nebula out to the shell. As recently shown by Pilyugin et
al. (\cite{temperature}), these line ratios can achieve very high
accuracies. It is also shown that the combination of those lines with
\ion{H}{$\beta$} can directly achieve abundances. Furthermore, a detailed
investigation of the high excitation features in the
inner hot wind bubble and of the wind itself emerging the CSPN is
suggested. It is desired to complete the whole image and to give input to
multi-dimensional radiative transfer hydrodynamic calculations.

\begin{acknowledgements}
      We would like to thank the (anonymous) referee and the editor Malcolm Walmsley for helping to improve the manuscript. We thank Thomas Rauch (T\"ubingen) for extensive discussions, help with the stellar atmosphere models and providing us the ESO~3.6m (taken under proposal 64.H-0557) and
      SAAO~1.9m data. We thank Greg
      Taylor (U. of New Mexico) for providing the original FITS image of his VLA observations.\newline
      S.{\"O}. is supported by the Austrian
      \emph{Fonds zur Wissenschaftlichen Forschung, FWF\/} doctoral school project
      W1227.\newline
      The HST Archival images were taken under Proposal 11827 (P.I. Keith Noll). This research has made use of the SIMBAD and Aladin data bases,
      operated at CDS, Strasbourg, France.
\end{acknowledgements}

%\newpage
%\phantom{XXX}
%\newpage
\onecolumn
\begin{appendix}

\section{The model results of the main nebula}
\label{model_list_appendix}
The table gives the full model of the best fit. The normalization on the left hand side is relative to \ion{H}{$\beta$} at each column individually. This allows the view on line ratios as they are commonly used in nebular spectroscopy. The right hand table is normalized to \ion{H}{$\beta$} only in the middle column, to give a better view on the radial evolution of individual line fluxes.
\begin{table*}[!ht]
\caption{The measured line intensities vs. the model lines of the best fit
  model of the main nebula. }
\label{model_list.tab}
\centering
\begin{tabular}{llcc cc cc}
\hline\hline
\multicolumn{2}{l}{line ($\lambda\,[$\AA$])$} & \multicolumn{2}{c}{$3.5\times10^{17}$cm} & \multicolumn{2}{c}{$6.5\times10^{17}$cm} &
\multicolumn{2}{c}{$9.0\times10^{17}$cm} \\
\multicolumn{2}{l}{ } & \multicolumn{2}{c}{$\equiv$13\arcsec} & \multicolumn{2}{c}{$\equiv$23\arcsec} &
\multicolumn{2}{c}{$\equiv$32\arcsec} \\
\multicolumn{2}{l}{} & obs. & model & obs. & model & obs. & model \\
\hline
\hline
\\
\vspace{3pt}
\parbox[c]{0.2cm}{
\ion{H}{$\theta$}
\ion{He}{ii}} &
$\!\!\!\!\!\!\left.{\parbox[c]{0.84cm}{
(3798)
(3797)}}\right\}$  &
6.7 &
$\!\!\left\{{\parbox[c]{0.45cm}{
5.57
0.11}}\right.$ &
7.0 &
$\!\!\left\{{\parbox[c]{0.45cm}{
5.35
0.04}}\right.$ &
6.2 &
$\!\!\left\{{\parbox[c]{0.45cm}{
5.3
--- }}\right.$
\\
\vspace{1pt}
\ion{H}{$\eta$}   & $\left.{\!\!\!\!\!\!\parbox[c]{0.84cm}{(3835)}}\right.$ &
9.6     & 7.7     & 9.7     & 7.4     & 7.2     & 7.3     \\
\vspace{1pt}
\ion{Ne}{iii}   & $\left.{\!\!\!\!\!\!\parbox[c]{0.84cm}{(3869)}}\right.$ &
122     & 140    & 140     & 116     & 170     & 138  \\
\vspace{3pt}
\parbox[c]{0.2cm}{
\ion{H}{$\zeta$}
\ion{He}{i}} &
$\!\!\!\!\!\!\left.{\parbox[c]{0.84cm}{
(3889)
(3889)}}\right\}$  &
22 &
$\!\!\left\{{\parbox[c]{0.45cm}{
10.9
10.4}}\right.$ &
23 &
$\!\!\left\{{\parbox[c]{0.45cm}{
10.5
\phantom{1}9.7}}\right.$ &
30 &
$\!\!\left\{{\parbox[c]{0.45cm}{
10.4
10.9}}\right.$
\\\vspace{3pt}
\parbox[c]{0.2cm}{
\ion{H}{$\epsilon$}
\ion{Ne}{iii}} &
$\!\!\!\!\!\!\left.{\parbox[c]{0.84cm}{
(3970)
(3968)}}\right\}$  &
52 &
$\!\!\left\{{
\parbox[c]{0.45cm}{
16.3
42.2}}\right.$
&
62 &
$\!\!\left\{{\parbox[c]{0.45cm}{
15.9
35.0}}\right.$ &
68 &
$\!\!\left\{{\parbox[c]{0.45cm}{
16.0
41.8}}\right.$
\\
\vspace{1pt}\ion{He}{i}   & $\left.{\!\!\!\!\!\!\parbox[c]{0.84cm}{(4026)}}\right.$ &
2.5     & 2.21     & 2.1     & 2.11     & 3.4     & 2.36     \\
\vspace{1pt}[\ion{S}{ii}]   & $\left.{\!\!\!\!\!\!\parbox[c]{0.84cm}{(4070)}}\right.$ &
2.4     & 3.76     & 3.9     & 3.98     & 8.3     & 5.62     \\
\vspace{1pt}\ion{H}{$\delta$}   & $\left.{\!\!\!\!\!\!\parbox[c]{0.84cm}{(4102)}}\right.$ &
27     & 26.4     & 26     & 25.9     & 26     & 26.0   \\
\vspace{1pt}\ion{He}{ii}   & $\left.{\!\!\!\!\!\!\parbox[c]{0.84cm}{(4200)}}\right.$ &
0.7    & 0.69     & 0.6     & 0.27     &      &     \\
\vspace{1pt}\ion{C}{ii}   & $\left.{\!\!\!\!\!\!\parbox[c]{0.84cm}{(4267)}}\right.$ &
1.8     & 0.66     & 1.0     & 0.69     & 0.9     & 0.67   \\
\vspace{1pt}\ion{H}{$\gamma$}   & $\left.{\!\!\!\!\!\!\parbox[c]{0.84cm}{(4340)}}\right.$ &
49    & 47.2     & 51     & 46.8     & 51     & 46.9   \\
\vspace{1pt}[\ion{O}{iii}]   & $\left.{\!\!\!\!\!\!\parbox[c]{0.84cm}{(4363)}}\right.$ &
10     & 9.77     & 10     & 6.13     & 8     & 6.56   \\
\vspace{1pt}\underline{\ion{He}{i}}   & $\left.{\!\!\!\!\!\!\parbox[c]{0.84cm}{(4471)}}\right.$ &
  4.6  & 4.62     & 4.0     & 4.39     & 6.8    &  4.93   \\
\vspace{1pt}\ion{He}{ii}   & $\left.{\!\!\!\!\!\!\parbox[c]{0.84cm}{(4542)}}\right.$ &
1.8    & 1.32     & 0.9     & 0.49     &      &     \\
\vspace{1pt}\underline{\ion{He}{ii}}   & $\left.{\!\!\!\!\!\!\parbox[c]{0.84cm}{(4686)}}\right.$ &
49    & 39.6     & 30     & 31.5     &  5    &  2.5   \\
\vspace{3pt}
\parbox[c]{0.2cm}{
\ion{He}{i}
\ion{Ar}{iv}
\ion{Ne}{iv}} &
$\!\!\!\!\!\!\left.{\parbox[c]{0.84cm}{
(4713)
(4711)
(4714)}}\right\}$ &
3.0 &
$\!\!\left\{{\parbox[c]{0.45cm}{
0.49
2.85
0.08 }}\right.$ &
1.5 &
$\!\!\left\{{\parbox[c]{0.45cm}{
0.45
0.64
---}}\right.$ &
&
$\!\!\left.{\parbox[c]{0.45cm}{

}}\right.$
\\
\vspace{1pt}\ion{Ar}{iV}   & $\left.{\!\!\!\!\!\!\parbox[c]{0.84cm}{(4740)}}\right.$ &
1.6    & 2.18     & 0.6     & 0.49     &      &     \\
\vspace{1pt}\ion{H}{$\beta$}   & $\left.{\!\!\!\!\!\!\parbox[c]{0.84cm}{(4961)}}\right.$ &
{\bf 100}    & {\bf 100}     & {\bf 100}     & {\bf 100}     &  {\bf 100}    &  {\bf 100}   \\
\vspace{1pt}\ion{He}{i}   & $\left.{\!\!\!\!\!\!\parbox[c]{0.84cm}{(4922)}}\right.$ &
1.1    & 1.27     & 1.1    & 1.22     &  1.4    &  1.36   \\
\vspace{1pt}[\ion{O}{iii}]   & $\left.{\!\!\!\!\!\!\parbox[c]{0.84cm}{(4959)}}\right.$ &
325    & 363     & 327     & 310     &  240    &  293   \\
\vspace{1pt}\underline{[\ion{O}{iii}]}   & $\left.{\!\!\!\!\!\!\parbox[c]{0.84cm}{(5007)}}\right.$ &
979    & 1097     & 982     & 935     &  723    &  882   \\
\vspace{1pt}\ion{N}{i}   & $\left.{\!\!\!\!\!\!\parbox[c]{0.84cm}{(5199)}}\right.$ &
1.6    & 0.81     & 2.8     & 0.84     &  13.2    &  2.02   \\
\vspace{1pt}\ion{Ne}{ii}   & $\left.{\!\!\!\!\!\!\parbox[c]{0.84cm}{(5412)}}\right.$ &
4.0    & 3.02     & 2.6     & 1.16     &       &      \\
\vspace{1pt}\ion{Cl}{iii}   & $\left.{\!\!\!\!\!\!\parbox[c]{0.84cm}{(5518)}}\right.$ &
     &       & 1.6     & 1.15     &       &      \\
\vspace{1pt}\ion{Cl}{iii}   & $\left.{\!\!\!\!\!\!\parbox[c]{0.84cm}{(5538)}}\right.$ &
     &       & 0.8     & 0.86     &       &      \\
\vspace{1pt}[\ion{N}{ii}]   & $\left.{\!\!\!\!\!\!\parbox[c]{0.84cm}{(5755)}}\right.$ &
  2.8   & 3.48      & 3.9     & 3.63     &  11.9     &  6.93    \\
\vspace{1pt}\ion{He}{i}   & $\left.{\!\!\!\!\!\!\parbox[c]{0.84cm}{(5876)}}\right.$ &
  13   & 13.42      & 13     & 12.95     &  23     &  13.64    \\
\vspace{1pt}[\ion{N}{ii}]   & $\left.{\!\!\!\!\!\!\parbox[c]{0.84cm}{(6548)}}\right.$ &
  64   & 81     & 93     & 87     &  170     &  151    \\
\vspace{1pt}\underline{\ion{H}{$\alpha$}}   & $\left.{\!\!\!\!\!\!\parbox[c]{0.84cm}{(6563)}}\right.$ &
  293   & 290      & 294     & 291     &  291     &  291    \\
\vspace{1pt}[\ion{N}{ii}]   & $\left.{\!\!\!\!\!\!\parbox[c]{0.84cm}{(6584)}}\right.$ &
  178   & 239      & 279     & 256     &  434     &  449    \\
\vspace{1pt}\ion{He}{i}   & $\left.{\!\!\!\!\!\!\parbox[c]{0.84cm}{(6678)}}\right.$ &
  2.4   & 3.80      & 4.3     & 3.67    &  5.7     &  4.04    \\
\vspace{1pt}[\ion{S}{ii}]   & $\left.{\!\!\!\!\!\!\parbox[c]{0.84cm}{(6716)}}\right.$ &
  19   & 29.3      & 37     & 31.44    &  74     &  50.0    \\
\vspace{1pt}[\ion{S}{ii}]   & $\left.{\!\!\!\!\!\!\parbox[c]{0.84cm}{(6731)}}\right.$ &
  15  & 26.1      & 31     & 28.2    &  66     &  40.0    \\
\vspace{1pt}\ion{He}{i}   & $\left.{\!\!\!\!\!\!\parbox[c]{0.84cm}{(7065)}}\right.$ &
  3.4   & 2.62      & 2.7     & 2.46    &  1.7     &  2.67    \\
\vspace{1pt}\ion{Ar}{iii}   & $\left.{\!\!\!\!\!\!\parbox[c]{0.84cm}{(7135)}}\right.$ &
  22   & 25.9      & 25     & 27.2    &  23     &  31.5    \\
\hline
\end{tabular}
%===================================================
\phantom{X}
\begin{tabular}{llcc cc cc}
\hline\hline
\multicolumn{2}{l}{line ($\lambda\,[$\AA$])$} & \multicolumn{2}{c}{$3.5\times10^{17}$cm} & \multicolumn{2}{c}{$6.5\times10^{17}$cm} &
\multicolumn{2}{c}{$9.0\times10^{17}$cm} \\
\multicolumn{2}{l}{ } & \multicolumn{2}{c}{$\equiv$13\arcsec} & \multicolumn{2}{c}{$\equiv$23\arcsec} &
\multicolumn{2}{c}{$\equiv$32\arcsec} \\
\multicolumn{2}{l}{} & obs. & model & obs. & model & obs. & model \\
\hline
\hline
\\
\vspace{3pt}
\parbox[c]{0.2cm}{
\ion{H}{$\theta$}
\ion{He}{ii}} &
$\!\!\!\!\!\!\left.{\parbox[c]{0.84cm}{
(3798)
(3797)}}\right\}$  &
4.5 &
$\!\!\left\{{\parbox[c]{0.45cm}{
3.63
0.07}}\right.$ &
6.9 &
$\!\!\left\{{\parbox[c]{0.45cm}{
5.35
0.04}}\right.$ &
2.9 &
$\!\!\left\{{\parbox[c]{0.45cm}{
2.4
--- }}\right.$
\\
\vspace{1pt}
\ion{H}{$\eta$}   & $\left.{\!\!\!\!\!\!\parbox[c]{0.84cm}{(3835)}}\right.$ &
6.4     & 5.0     & 9.5     & 7.4     & 3.4     & 3.3     \\
\vspace{1pt}
\ion{Ne}{iii}   & $\left.{\!\!\!\!\!\!\parbox[c]{0.84cm}{(3869)}}\right.$ &
82     & 91     & 140     & 116     & 80     & 62  \\
\vspace{3pt}
\parbox[c]{0.2cm}{
\ion{H}{$\zeta$}
\ion{He}{i}} &
$\!\!\!\!\!\!\left.{\parbox[c]{0.84cm}{
(3889)
(3889)}}\right\}$  &
15 &
$\!\!\left\{{\parbox[c]{0.45cm}{
7.1
6.8}}\right.$ &
23 &
$\!\!\left\{{\parbox[c]{0.45cm}{
10.5
\phantom{1}9.7}}\right.$ &
14 &
$\!\!\left\{{\parbox[c]{0.45cm}{
4.7
4.9}}\right.$
\\\vspace{3pt}
\parbox[c]{0.2cm}{
\ion{H}{$\epsilon$}
\ion{Ne}{iii}} &
$\!\!\!\!\!\!\left.{\parbox[c]{0.84cm}{
(3970)
(3968)}}\right\}$  &
35 &
$\!\!\left\{{
\parbox[c]{0.45cm}{
10.6
27.5}}\right.$
&
61 &
$\!\!\left\{{\parbox[c]{0.45cm}{
15.9
35.0}}\right.$ &
32 &
$\!\!\left\{{\parbox[c]{0.45cm}{
~\,\,7.2
18.8}}\right.$
\\
\vspace{1pt}\ion{He}{i}   & $\left.{\!\!\!\!\!\!\parbox[c]{0.84cm}{(4026)}}\right.$ &
1.7     & 1.44     & 2.1     & 2.11     & 1.6     & 1.06     \\
\vspace{1pt}[\ion{S}{ii}]   & $\left.{\!\!\!\!\!\!\parbox[c]{0.84cm}{(4070)}}\right.$ &
1.6     & 2.45     & 3.8     & 3.98     & 3.9     & 2.53     \\
\vspace{1pt}\ion{H}{$\delta$}   & $\left.{\!\!\!\!\!\!\parbox[c]{0.84cm}{(4102)}}\right.$ &
18     & 17.2     & 25     & 25.9     & 12     & 11.7   \\
\vspace{1pt}\ion{He}{ii}   & $\left.{\!\!\!\!\!\!\parbox[c]{0.84cm}{(4200)}}\right.$ &
0.5    & 0.45     & 0.6     & 0.27     &      &     \\
\vspace{1pt}\ion{C}{ii}   & $\left.{\!\!\!\!\!\!\parbox[c]{0.84cm}{(4267)}}\right.$ &
1.2     & 0.43     & 1.0     & 0.69     & 0.4     & 0.30   \\
\vspace{1pt}\ion{H}{$\gamma$}   & $\left.{\!\!\!\!\!\!\parbox[c]{0.84cm}{(4340)}}\right.$ &
33    & 30.8     & 50     & 46.8     & 24     & 21.1   \\
\vspace{1pt}[\ion{O}{iii}]   & $\left.{\!\!\!\!\!\!\parbox[c]{0.84cm}{(4363)}}\right.$ &
6.7     & 6.37     & 9.8     & 6.13     & 3.7     & 2.95   \\
\vspace{1pt}\underline{\ion{He}{i}}   & $\left.{\!\!\!\!\!\!\parbox[c]{0.84cm}{(4471)}}\right.$ &
  3.1  & 3.01     & 3.9     & 4.39     &  3.2    &  2.22   \\
\vspace{1pt}\ion{He}{ii}   & $\left.{\!\!\!\!\!\!\parbox[c]{0.84cm}{(4542)}}\right.$ &
1.2    & 0.86     & 0.9     & 0.49     &      &     \\
\vspace{1pt}\underline{\ion{He}{ii}}   & $\left.{\!\!\!\!\!\!\parbox[c]{0.84cm}{(4686)}}\right.$ &
33    & 25.8     & 29     & 31.5     &  2.3    &  1.14   \\
\vspace{3pt}
\parbox[c]{0.2cm}{
\ion{He}{i}
\ion{Ar}{iv}
\ion{Ne}{iv}} &
$\!\!\!\!\!\!\left.{\parbox[c]{0.84cm}{
(4713)
(4711)
(4714)}}\right\}$ &
2.0 &
$\!\!\left\{{\parbox[c]{0.45cm}{
0.32
1.86
0.05 }}\right.$ &
1.44 &
$\!\!\left\{{\parbox[c]{0.45cm}{
0.45
0.64
---}}\right.$ &
&
$\!\!\left.{\parbox[c]{0.45cm}{

}}\right.$
\\
\vspace{1pt}\ion{Ar}{iV}   & $\left.{\!\!\!\!\!\!\parbox[c]{0.84cm}{(4740)}}\right.$ &
1.1    & 1.42     & 0.6     & 0.49     &      &     \\
\vspace{1pt}\ion{H}{$\beta$}   & $\left.{\!\!\!\!\!\!\parbox[c]{0.84cm}{(4961)}}\right.$ &
67    & 65.2     & 98     & {\bf 100}     &  47    &  45   \\
\vspace{1pt}\ion{He}{i}   & $\left.{\!\!\!\!\!\!\parbox[c]{0.84cm}{(4922)}}\right.$ &
0.75    & 0.83     & 1.1    & 1.22     &  0.66    &  0.61   \\
\vspace{1pt}[\ion{O}{iii}]   & $\left.{\!\!\!\!\!\!\parbox[c]{0.84cm}{(4959)}}\right.$ &
218    & 237     & 320     & 310     &  113    &  132   \\
\vspace{1pt}\underline{[\ion{O}{iii}]}   & $\left.{\!\!\!\!\!\!\parbox[c]{0.84cm}{(5007)}}\right.$ &
656    & 715     & 962     & 935     &  340    &  397   \\
\vspace{1pt}\ion{N}{i}   & $\left.{\!\!\!\!\!\!\parbox[c]{0.84cm}{(5199)}}\right.$ &
1.1    & 0.53     & 2.7     & 0.84     &  6.2    &  0.91   \\
\vspace{1pt}\ion{Ne}{ii}   & $\left.{\!\!\!\!\!\!\parbox[c]{0.84cm}{(5412)}}\right.$ &
2.7    & 1.97     & 2.5     & 1.16     &       &      \\
\vspace{1pt}\ion{Cl}{iii}   & $\left.{\!\!\!\!\!\!\parbox[c]{0.84cm}{(5518)}}\right.$ &
     &       & 1.6     & 1.15     &       &      \\
\vspace{1pt}\ion{Cl}{iii}   & $\left.{\!\!\!\!\!\!\parbox[c]{0.84cm}{(5538)}}\right.$ &
     &       & 0.8     & 0.86     &       &      \\
\vspace{1pt}[\ion{N}{ii}]   & $\left.{\!\!\!\!\!\!\parbox[c]{0.84cm}{(5755)}}\right.$ &
  1.9   & 2.27      & 3.9     & 3.63     &  5.6     &  3.12    \\
\vspace{1pt}\ion{He}{i}   & $\left.{\!\!\!\!\!\!\parbox[c]{0.84cm}{(5876)}}\right.$ &
  9.0   & 8.75      & 12.6     & 12.95     &  11.2     &  6.14    \\
\vspace{1pt}[\ion{N}{ii}]   & $\left.{\!\!\!\!\!\!\parbox[c]{0.84cm}{(6548)}}\right.$ &
  43   & 53     & 91     & 87     &  80     &  68    \\
\vspace{1pt}\underline{\ion{H}{$\alpha$}}   & $\left.{\!\!\!\!\!\!\parbox[c]{0.84cm}{(6563)}}\right.$ &
  203   & 189      & 290     & 291     &  137     &  131    \\
\vspace{1pt}[\ion{N}{ii}]   & $\left.{\!\!\!\!\!\!\parbox[c]{0.84cm}{(6584)}}\right.$ &
  119   & 156      & 274     & 256     &  204     &  202    \\
\vspace{1pt}\ion{He}{i}   & $\left.{\!\!\!\!\!\!\parbox[c]{0.84cm}{(6678)}}\right.$ &
  1.6   & 2.48      & 4.2     & 3.67    &  2.7     &  1.82    \\
\vspace{1pt}[\ion{S}{ii}]   & $\left.{\!\!\!\!\!\!\parbox[c]{0.84cm}{(6716)}}\right.$ &
  13   & 19.1      & 36     & 31.4    &  35     &  22.4    \\
\vspace{1pt}[\ion{S}{ii}]   & $\left.{\!\!\!\!\!\!\parbox[c]{0.84cm}{(6731)}}\right.$ &
  10   & 17.0      & 30     & 28.2    &  31     &  17.9    \\
\vspace{1pt}\ion{He}{i}   & $\left.{\!\!\!\!\!\!\parbox[c]{0.84cm}{(7065)}}\right.$ &
  2.3   & 1.71      & 2.6     & 2.46    &  0.8     &  1.20    \\
\vspace{1pt}\ion{Ar}{iii}   & $\left.{\!\!\!\!\!\!\parbox[c]{0.84cm}{(7135)}}\right.$ &
  15   & 16.9     & 25     & 27.2    &  11     &  14.2    \\
\hline
\end{tabular}
\tablefoot{Radii correspond to those described in Tab.\,\ref{model_input.tab}. The four lines used for the  fitting procedure of the free parameters
  are underlined. The lines are normalized to \ion{H}{$\beta$} = 100 at each column individually in the left part of the table (marked in bold face). The right hand part gives the fluxes relative to
  the brightest point in \ion{H}{$\beta$} = 100 ($\equiv\,\,2\,\times\,10^{-13}\,{\rm ergs}\,\,{\rm cm}^{-1}\,\,
  {\rm s}^{-1}\,\,${\AA}$^{-1}$ per slit). The regions are selected as {\sl 'inner edge'}, {\sl '\ion{H}{$\alpha$} peak'} and {\sl 'outer edge'}. Each of them is a narrow screenshot of only 2\arcsec~.}
\end{table*}

\end{appendix}
\end{document}